\documentclass[11pt,oneside,english]{article}

\usepackage{amsmath,amsfonts,amssymb}
\usepackage{authblk}
\usepackage{bm}
\usepackage{booktabs}
\usepackage[font=footnotesize]{caption}
\usepackage{cite}
\usepackage[T1]{fontenc}
\usepackage[verbose,tmargin=2.6cm,bmargin=2.6cm,lmargin=2.5cm,rmargin=2.5cm,footskip=1cm]{geometry}
\usepackage[utf8]{inputenc}
\usepackage{lmodern}
\usepackage{mathrsfs}
\usepackage{mathtools}
\usepackage{xcolor}
\usepackage[colorlinks=true,linkcolor={red!70!black},citecolor={blue!90},urlcolor={blue!80!black}]{hyperref}
\usepackage{tensor}

\usepackage{mmacells}
\mmaSet{
  morefv={gobble=7}
}
\mmaDefineMathReplacement{𝕖}{\mathbbm{e}}
\mmaDefineMathReplacement{`}{\text{\`{}}}

\usepackage{tikz}
\usetikzlibrary{arrows,decorations.markings}
\tikzset{
  big arrow/.style={
    decoration={markings,mark=at position 1 with {\arrow[scale=2,#1]{>}}},
    postaction={decorate},
    shorten >=0.4pt},
  big arrow/.default=black}

\usepackage[nameinlink,capitalize]{cleveref}
\crefname{figure}{Figure}{Figures}
\labelcrefformat{subequation}{#2(#1)#3}
\labelcrefrangeformat{subequation}{#3(#1)#4 to #5(#2)#6}

\numberwithin{equation}{section}

\makeatletter
\renewenvironment{abstract}{%
    \if@twocolumn
      \section*{\abstractname}%
    \else 
      \begin{center}%
        {\bfseries \normalsize\abstractname\vspace{\z@}}
      \end{center}%
      \quotation
    \fi}
    {\if@twocolumn\else\endquotation\fi}
\makeatother

\makeatletter
\g@addto@macro\bfseries{\boldmath}

\let\originalleft\left
\let\originalright\right
\renewcommand*{\left}{\mathopen{}\mathclose\bgroup\originalleft}
\renewcommand*{\right}{\aftergroup\egroup\originalright}

\renewcommand*{\to}{\mathchoice{\longrightarrow}{\rightarrow}{\rightarrow}{\rightarrow}}

\DeclareMathOperator{\SU}{SU}
\DeclareMathOperator{\SL}{SL}
\DeclareMathOperator{\SO}{SO}
\DeclareMathOperator{\Sp}{Sp}
\DeclareMathOperator{\gA}{A}
\DeclareMathOperator{\gB}{B}
\DeclareMathOperator{\gC}{C}
\DeclareMathOperator{\gD}{D}
\DeclareMathOperator{\gE}{E}
\DeclareMathOperator{\gF}{F}
\DeclareMathOperator{\gG}{G}

\newcommand*{\G}{\ensuremath G}

\DeclareMathOperator{\Lie}{Lie}
\DeclareMathOperator{\rank}{rank}
\DeclareMathOperator{\ord}{ord}
\DeclareMathOperator{\PD}{PD}
\DeclareMathOperator{\vol}{vol}

\newcommand*{\singtype}[1]{\ensuremath \text{#1}}

\newcommand*{\bP}{\ensuremath\mathbb{P}}
\newcommand*{\bR}{\ensuremath\mathbb{R}}
\newcommand*{\bZ}{\ensuremath\mathbb{Z}}
\newcommand*{\cF}{\ensuremath\mathcal{F}}
\newcommand*{\cN}{\ensuremath\mathcal{N}}
\newcommand*{\cW}{\ensuremath\mathcal{W}}
\newcommand*{\cZ}{\ensuremath\mathcal{Z}}
\newcommand*{\sfr}{\ensuremath\mathsf{r}}
\newcommand*{\sK}{\ensuremath\mathcal{K}}
\newcommand*{\sL}{\ensuremath\mathcal{L}}
\newcommand*{\sO}{\ensuremath\mathcal{O}}
\newcommand*{\sV}{\ensuremath\mathcal{V}}

\DeclarePairedDelimiter{\ceil}{\lceil}{\rceil}
\DeclarePairedDelimiter{\floor}{\lfloor}{\rfloor}

\makeatletter
\newcommand*{\diff}{\@ifnextchar^{\DIFF}{\DIFF^{}}}
\def\DIFF^#1{\mathop{\mathrm{\mathstrut d}}\nolimits^{#1}\gobblespace}
\def\gobblespace{\futurelet\diffarg\opspace}
\def\opspace{%
    \let\DiffSpace\!%
    \ifx\diffarg(%
        \let\DiffSpace\relax
    \else
        \ifx\diffarg[%
            \let\DiffSpace\relax
        \else
            \ifx\diffarg\{%
                \let\DiffSpace\relax
            \fi
        \fi
    \fi
    \DiffSpace
}
\makeatother

\title{
\Huge Generating functions for intersection products of divisors in resolved F-theory models}
\author{\Large Patrick Jefferson\thanks{pjeffers@mit.edu}}
\affil{\normalsize \emph{Center for Theoretical Physics, Department of Physics, Massachusetts Institute of Technology, 77 Massachusetts Avenue, Cambridge, MA 02139, USA}}
\author{\Large Andrew P. Turner\thanks{turnerap@sas.upenn.edu}}
\affil{\normalsize \emph{Department of Physics and Astronomy, University of Pennsylvania, Philadelphia, PA 19104, USA}}

\date{}

\begin{document}
\maketitle
\begin{tikzpicture}[remember picture,overlay]
   \node[anchor=north east,inner sep=0pt] at (current page.north east)
              {$\begin{array}{ccc}&&\\ \\ \text{MIT-CTP-5424}&&\end{array}$};
\end{tikzpicture}
\thispagestyle{empty}
\begin{abstract}
\noindent Building on the approach of \texttt{1703.00905}, we present an efficient algorithm for computing topological intersection numbers of divisors in a broad class of elliptic fibrations with the aid of a symbolic computing tool. A key part of our strategy is organizing the intersection products of divisors into a succinct analytic generating function, namely the exponential of the K\"ahler class. We use the methods of \texttt{1703.00905} to compute the pushforward of this function to the base of the elliptic fibration. We implement our algorithm in an accompanying \emph{Mathematica} package \texttt{IntersectionNumbers.m} that computes generating functions of intersection products for resolutions of F-theory Tate models defined over smooth base of arbitrary complex dimension. Our algorithm appears to offer a significant reduction in computation time needed to compute intersection numbers as compared to previously explored implementations of the methods in \texttt{1703.00905}; as an illustration, we explicitly compute the generating functions for all F-theory Tate models with simple classical groups of rank up to twenty and highlight the growth of the computation time with the rank of the group.
\end{abstract}
\flushbottom
\newpage
\tableofcontents
\addtocontents{toc}{\protect\thispagestyle{empty}}
\setcounter{page}{1}
\section{Introduction}
\label{sec:intro}

Calabi--Yau (CY) $n$-folds, Ricci-flat K\"ahler manifolds of complex dimension $n$ with global\footnote{Various closely related, but inequivalent definitions of CY manifolds appear throughout the literature (see, e.g.,~\cite{TianYau1,TianYau2,GHJ}). Although a common definition defines a CY $n$-fold as a Ricci-flat K\"ahler manifold of complex dimension $n$, this is somewhat insufficient for string theory applications as Ricci-flatness only implies local (i.e., not necessarily global) $\SU(n)$ holonomy; we therefore adopt a more restrictive definition to ensure global $\SU(n)$ holonomy.} holonomy a subgroup of $\SU(n)$, play a central role in string theory compactifications. Ricci-flatness guarantees that these manifolds are vacuum solutions of the Einstein field equations and special holonomy implies that a fraction of supersymmetry, depending on the precise holonomy group, can be preserved in the lower dimensional theory~\cite{JoyceCompact}. A choice of string duality frame and CY manifold determine the dimension and manifest supersymmetry of the resulting effective theory describing the compactification at low energies, and moreover there is detailed correspondence between the geometric properties of the CY compactification space and the physical properties of the vacuum solutions of the effective theory.

In most string duality frames, compactifications on a generic choice of CY manifold are expected to produce consistent effective theories. A notable exception is F-theory~\cite{Vafa:1996xn,Morrison:1996na,Morrison:1996pp}, a generically non-perturbative twelve-dimensional geometric formulation of type IIB string theory vacua in which the extra two dimensions (relative to the critical ten dimensions of superstring theory) are required to take the shape of a complex torus\footnote{The overall volume of the torus is unphysical in the F-theory duality frame.}. This unusual property of F-theory vacuum spacetimes implies that F-theory compactifications can only be formulated consistently on a special subset of CY $n$-folds that exhibit the structure of a genus one fibration\footnote{A genus one fibration has fibers isomorphic to genus one complex curves.} over a K\"ahler manifold of complex dimension $n - 1$. F-theory vacua are therefore generally characterized by elliptically fibered\footnote{An elliptic curve is a genus one complex curve with a choice rational point. Although the existence of a rational point is not essential to the self-consistency of F-theory~\cite{Braun:2014oya}, in this paper we restrict our attention to elliptically fibered CY $n$-folds for simplicity. The results of this paper can be straightforwardly adapted to the case of genus one-fibered CY $n$-folds. \label{foot:genusone}} CY $n$-folds.

Remarkably, F-theory vacua can accommodate nonperturbative configurations of backreacted 7-branes and O7-planes by permitting singular geometries, in which the elliptic fiber degenerates over special loci in the base of the fibration. Elliptic fiber singularities indicate the presence of massless (or tensionless) gauge-charged degrees of freedom in the low-energy spectrum; these singularities must be lifted in order to obtain a conventional low-energy effective description of the F-theory vacuum in terms of a local action. The lifting of these light BPS degrees of freedom can be accomplished by compactifying the low-energy theory on a circle and switching on non-trivial VEVs for massless vector multiplet scalars\footnote{These VEVs are the holonomies of gauge fields integrated over the compact circle.} appearing in the action of the circle-compactified theory. Switching on these VEVs Higgses the theory and gives a nonzero mass or tension to light degrees of freedom in the low-energy spectrum; this can be viewed as deforming the theory towards a generic locus of the Coulomb branch of the moduli space.\footnote{Note that these moduli may be lifted in compactifications preserving four or fewer supercharges.} Geometrically, these deformations correspond to K\"ahler deformations of the CY background that lead to related but topologically distinct geometries in which the volumes of the elliptic fibers (and hence, the calibrated volumes of all holomorphic cycles) are strictly positive. The resulting smooth elliptic CY $n$-fold is referred to as a resolution of the original singular CY space.

Resolving the singularities of the compactification space in a manner that preserves the CY condition realizes the strongly-coupled physics associated to the singularities as a degenerate limit of a smooth geometric background. In order to interpret the physics of the resolution, one can use the fact that F-theory compactified on a smooth CY $n$-fold times a transverse circle is dual to M-theory compactified on the same CY $n$-fold; here, the radius of the Kaluza--Klein (KK) circle in the F-theory frame is proportional to the inverse volume of the elliptic fiber in the M-theory frame. The resulting effective KK theory describing the M-theory compactification at low energies is defined by a supersymmetric action whose kinematics can be computed directly from the geometry of the smooth CY background using classical techniques in complex and algebraic geometry. The dimensional uplift of this KK theory, i.e., the limit in which the KK radius becomes infinitely large, is then understood in the M-theory frame to correspond to the limit in which the volume of elliptic fiber goes to zero, forcing the K\"ahler deformations responsible for resolving the singularities to vanish, and driving the geometry back to the special locus in K\"ahler moduli space corresponding to the singular F-theory limit.

The purpose of this paper is to describe an efficient algorithm for computing a specific set of topological invariants of smooth elliptically fibered CY $n$-folds $X^{(n)}$ resolving singular F-theory backgrounds, namely the $n$-fold intersection numbers of divisors\footnote{A divisor $\hat{D}_I$ of a smooth K\"ahler $n$-fold is a holomorphic cycle of complex codimension one, i.e., a holomorphic $(n - 1)$-cycle. As we discuss in \cref{sec:pushforwards}, a smooth CY $n$-fold $X^{(n)}$ admits a basis of $h^{1, 1}(X^{(n)})$ divisors $\hat{D}_I$ that are Poincar\'e dual to harmonic $(1, 1)$-forms $\omega_I$. Note that Lefshetz's theorem on $(1, 1)$-classes applied to projective varieties such as $X^{(n)}$ guarantees that given a basis of divisors $\hat{D}_I$, there always exists a corresponding Poincar\'e dual basis of harmonic $(1, 1)$ forms $\hat{\omega}_I \equiv \PD(\hat{D}_I)$, see, e.g.,~\cite{Bizet:2014uua}.}:
    \begin{equation}
    \label{basicint}
         \hat{D}_{I_1} \hat{D}_{I_2} \dotsm \hat{D}_{I_n} = \int_{X^{(n)}} \hat{\omega}_{I_1} \wedge \hat{\omega}_{I_2} \wedge \dotsb \wedge \hat{\omega}_{I_n}\,.
    \end{equation}
The intersection numbers of a smooth (or possibly resolved) elliptically fibered CY $n$-fold determine one-loop (perturbatively) exact couplings of the $(11 - 2 n)$D theory describing the dual M-theory compactification at low energies, where the precise correspondence between intersection numbers and low-energy couplings depends on the complex dimension $n$ and holonomy of the elliptic CY. Since the smooth elliptic CY $n$-fold is assumed to be a resolution of a singular CY space defining an F-theory vacuum, the low-energy $(11 - 2 n)$D theory is a KK-reduction of a theory living in one dimension higher and hence inherits the kinematic structure of its $(12 - 2 n)$D lift. In fact, it turns out that the one-loop couplings of the $(11 - 2 n)$D theory essentially capture the kinematics of the original $(12 - 2 n)$D theory and can be used to study various properties that are difficult to access due to the generically singular nature of F-theory vacua; see~\cite{Weigand:2018rez} for an excellent overview. Below, we describe some specific examples of F-theory compactifications and ways in which the intersection numbers of divisors can be used to determine properties of the corresponding low-energy $(12 - 2 n)$D theory and its KK-reduction.

\paragraph{Elliptic CY 3-folds} F-theory on a singular elliptic CY 3-fold is described at low energies by 6D $\cN = (1, 0)$ supergravity (i.e., eight supercharges), where the light degrees of freedom are conventionally described by a gravity multiplet and a collection of tensor, vector, and hypermultiplets whose charges and multiplicities are required to satisfy 6D anomaly cancellation~\cite{Sagnotti:1992qw,Erler:1993zy,Taylor:2011wt,Park:2011ji}. Theories with eight conserved supercharges, especially those in greater than four spacetime dimensions, are highly constrained by supersymmetry (see, e.g.,~\cite{Seiberg:1996bd,Seiberg:1996vs}), and it is possible to compute many features of their vacua purely in terms of the geometry of the compactification space~\cite{Witten:1996qb,Morrison:1996xf,Douglas:1996xp,Intriligator:1997pq,Diaconescu:1998cn}.

The feature most relevant to this discussion is the collection of vector multiplet moduli of the low-energy theory compactified on a circle, which correspond~\cite{Witten:1996qb} to K\"ahler moduli of an elliptic CY 3-fold. There is an elegant correspondence between the resolutions of a singular elliptic CY 3-fold and the Coulomb branch phases of the associated 6D $\cN = (1, 0)$ supergravity theory compactified on a circle. The one-loop exact quadratic and cubic couplings of the low-energy effective 5D KK theory associated to each phase can be determined geometrically by computing intersection numbers of divisors in the CY 3-fold. This correspondence has been checked in numerous examples, for instance by recovering information about the 6D spectrum (e.g., the anomaly cancellation conditions) from the triple intersection numbers of divisors~\cite{Bonetti:2011mw,Grimm:2013oga,Grimm:2015zea,Esole:2015xfa,Esole:2017qeh,Esole:2017hlw,Bies:2017abs,Esole:2018vnm,Corvilain:2020tfb}:
    \begin{equation}
    \begin{aligned}
        S_\text{5D} &= \dotsb +  \frac{1}{24 \pi^2}  k_{i j k} \int A^i \wedge \diff A^j \wedge \diff A^k + \dotsb\,, \qquad k_{i j k} = \int_{X^{(3)}} \hat{\omega}_i  \wedge \hat{\omega}_j \wedge \hat{\omega}_k\,.
    \end{aligned}
    \end{equation}

The robustness of the correspondence between resolutions of elliptic CY 3-fold singularities and 5D vector multiplet moduli spaces suggests an approach for testing whether or not a candidate 5D $\cN = 1$ supergravity theory admits a 6D lift with a UV completion in F-theory, simply by looking for the hallmarks of an elliptic fibration structure. For example, as discussed in~\cite{Huang:2018esr}, there is a conjecture due to Koll\'ar~\cite{Kollar:2012pv} asserting that a smooth CY $n$-fold $X^{(n)}$ is genus one fibered if and only if there exists a nef divisor $\hat{D}$ satisfying $\hat{D}^n = 0, \hat{D}^{n-1} \ne 0$, and this conjecture was in fact proven by Oguiso and Wilson~\cite{Oguiso,Wilson} for the special case $n = 3$. These geometric hallmarks have significantly shaped efforts to ``engineer'' UV complete 5D $\cN = 1$ KK theories using the intersection numbers of holomorphic surfaces, see, e.g.,~\cite{Jefferson:2018irk,Bhardwaj:2019fzv}.

Six- and five-dimensional theories without gravitational interactions in particular have received a great deal of interest recently, as they can be realized in string theory as local elliptic CY singularities and hence can be modeled by a broad class of local geometric constructions without reference to global topology. The intersection numbers of elliptic CY singularities have been used to geometrically classify 6D superconformal field theories compactified on a circle, whose mass deformations trigger RG flows to 5D superconformal fixed points, in terms of their one-loop couplings~\cite{Bhardwaj:2018yhy,Bhardwaj:2018vuu,Bhardwaj:2019fzv,Apruzzi:2019vpe,Apruzzi:2019opn,Apruzzi:2019enx,Apruzzi:2019kgb} (see also~\cite{Kim:2020hhh}). From a ``bottom up'' perspective, the hallmarks of an elliptic fibration structure in the quadratic and cubic couplings of a low-energy effective 5D KK theory can also be used to infer whether or not the theory admits a UV completion as a 6D fixed point theory constructed in F-theory~\cite{Jefferson:2017ahm,Bhardwaj:2020gyu}.

Aspects of CY geometry have also been used to devise conjectural tests for determining whether or not a low-energy effective theory (of fixed dimension) with gravitational dynamics admits a UV completion in a consistent theory of quantum gravity. The general idea behind these tests is that certain kinematic properties of low-energy effective theories with UV completions in string theory are in fact universal features of quantum gravity, and that effective theories failing to satisfy these constraints belong to the ``swampland'' of effective theories without UV completions in quantum gravity (see~\cite{Palti:2019pca} for a review).\footnote{An interesting implication of the swampland paradigm is that certain low-energy effective theories without gravitational dynamics cannot be consistently coupled to gravity (assuming a fixed number of spacetime dimensions), even if these non-gravitational theories are by themselves UV complete. Related questions were explored in the context of 6D F-theory compactifications in, e.g.,~\cite{DelZotto:2014fia,Hayashi:2019fsa}. From a geometric standpoint, the assertion that certain low-energy effective theories cannot be consistently coupled to gravity implies that the local CY constructions used to geometrically engineer these theories in string theory cannot describe a local ``patch'' of any compact CY that could be used to define a gravitational theory in the same number of dimensions.} The triple intersection numbers of CY 3-folds have played key role in determining specific swampland criteria believed to constrain the landscape of 5D supergravity theories; these constraints and their connection to 5D KK supergravity theories with lifts to 6D supergravity, which are expected to correspond to M-theory compactifications on \emph{elliptic} CY 3-folds, are described in~\cite{Katz:2020ewz}.\footnote{The fact that a low-energy effective theory coupled to gravity in a fixed number of dimensions may not admit a UV completion in string theory does not preclude the possibility that the low-energy effective theory could be consistently coupled to gravity in higher dimensions. This suggests that even though certain 5D $\cN = 1$ supergravity theories appear to lie in the 5D swampland, it may nonetheless be possible to couple these theories to gravity in $D \ge 6$. The 6D case is especially interesting, because elliptic CY 3-folds typically have multiple local patches that could be used to define 5D $\cN = 1$ gauge theories, and it may not be immediately obvious why these local patches cannot be embedded in non-elliptic compact CY 3-folds defining 5D $\cN = 1$ gravitational theories. In a certain sense, this obstruction is tied to the elliptic fibration, whose presence signals the existence of a tower of KK modes. Thus, in order to use an elliptic CY 3-fold to define a 5D supergravity theory, it is necessary to deform the geometry in a manner that removes the elliptic fibration and hence truncates the KK modes. Interestingly, and consistent with the observation in~\cite{Huang:2018esr} that the majority of known CY 3-folds admit at least one elliptic fibration, concrete examples of non-elliptic compact CY 3-folds are comparatively difficult to construct. We thank H.-C. Tarazi for discussions related to this point.}

\paragraph{Elliptic CY 4-folds} F-theory compactified on an elliptic CY 4-fold (see~\cite{Denef:2008wq} for a comprehensive overview) is described at low energies by 4D $\cN = 1$ supergravity (i.e., four supercharges), whose light degrees of freedom consist of a gravity multiplet and some number of vector and chiral multiplets whose multiplicities are required to satisfy anomaly cancellation~\cite{Grimm:2010ks,Grimm:2012yq}. Compared to theories with eight supercharges, theories with four conserved supercharges are far less constrained by supersymmetry. In particular, 4D $\cN = 1$ theories constructed in string theory have non-trivial K\"ahler potentials and superpotentials that typically exhibit complicated dependence on the geometry of the compactification space and background fluxes characterizing the string vacuum~\cite{Blumenhagen:2006ci,Douglas:2006es}. Consequently, it is highly non-trivial task to identify stable 4D $\cN = 1$ supersymmetric vacua in these setups. 


Despite these challenges, it is still possible to compute certain kinematic properties in either the limit of large K\"ahler moduli (i.e. the ``large volume limit'') and or the limit of large complex structure moduli (i.e. the ``large complex structure limit''), both of which considerably simplify certain aspects of the relationship between the geometry of the background elliptic CY 4-fold and the physics of the corresponding low energy effective 4D theory. Two kinematic properties of 4D $\cN=1$ theories relevant to the present discussion that do not depend on K\"ahler moduli, in particular, and can hence be computed in the large volume limit, are the massless chiral spectrum and the superpotential \cite{Beasley:2008dc}.

In 4D $\cN = 1$ gauge theories constructed from topologically twisted compactifications of higher dimensional theories (such as F-theory compactified on an elliptic CY 4-fold, which can be modeled locally as a topologically twisted compactification of the world-volume theory on a stack of 7-branes), the chirality of the matter spectrum is captured by a supersymmetry-protected quantity called the chiral index, which is equal to the number of chiral multiplets minus the number of anti-chiral multiplets. One can switch on a magnetic gauge flux background along the compact 7-brane world-volume directions to produce a non-trivial chiral index, which can then be computed by counting fermion zero modes (with signs) of a suitable choice of Dirac operator via the Atiyah--Singer index theorem~\cite{Beasley:2008dc,Beasley:2008kw,Donagi:2008ca}.

The chiral index is in fact a topological invariant of the compactification space and admits an interpretation as the integral of the Chern character of a line bundle times the Todd class of the tangent bundle of the compactification space defining the local 4D $\cN = 1$ theory. However, despite the fact that the chiral index admits an elegant geometric interpretation, F-theory theory in typical fashion encodes the chiral index in the singularities of an elliptic CY variety, making direct computation challenging. As usual, the standard approach to dealing with these singularities is to compactify the theory on a circle and switch on VEVs for 3D vector multiplet scalars corresponding to gauge holonomies around the KK circle in order to resolve the singularities. Analogous to the elliptic CY 3-fold case described above, the resolutions of an elliptic CY 4-fold correspond to the Coulomb branch\footnote{In 4D $\cN = 1$ theories, it is possible for quantum corrections to lift Coulomb branch directions, so the usage of the terminology ``Coulomb branch'' is provisional in this context.} phases of the 4D $\cN = 1$ theory compactified on a circle~\cite{Intriligator:2012ue}. Under M-theory/F-theory duality, an F-theory 7-brane worldvolume flux background is mapped to a background profile for the field strength of the M-theory 3-form, and the chiral indices $\chi_\sfr$ can then be recovered from the integrals of the flux background over certain algebraic 4-cycles in the resolved elliptic CY 4-fold~\cite{Marsano:2011hv,Borchmann:2013hta,Bies:2017fam}:
    \begin{equation}
        \chi_\sfr = \int_{S_\sfr} G_4 = \int_{X^{(4)}} \PD(S_\sfr) \wedge G_4\,.
    \end{equation}
These fluxes, at least in the case of ``vertical fluxes'', can be computed in terms of the quadruple intersection numbers of divisors in the resolved elliptic CY 4-fold,
    \begin{equation}
        \chi_\sfr = \chi_\sfr^{i j} \Theta_{i j}\,, \quad \Theta_{i j} = \int_{X^{(4)}} G_4 \wedge \hat{\omega}_i \wedge \hat{\omega}_j = G_4^{k l} \int_{X^{(4)}} \hat{\omega}_i \wedge \hat{\omega}_j \wedge \hat{\omega}_k \wedge \hat{\omega}_l\,,
    \end{equation}
and are visible in the 3D KK theory as one-loop (perturbatively) exact Chern--Simons couplings~\cite{Grimm:2011fx,Grimm:2011sk,Cvetic:2012xn,Cvetic:2013uta}:
    \begin{equation}
        S_\text{3D} = \dotsb + \Theta_{i j} \int A^i \wedge \diff A^j + \dotsb\,.
    \end{equation}
The correspondence between elliptic CY 4-fold geometry and the multiplicities of chiral matter has been used to explore the landscape of 4D $\cN = 1$ chiral theories~\cite{Marsano:2010ix,Grimm:2011tb,Braun:2011zm,Marsano:2011hv,Krause:2011xj,Grimm:2011fx,Jefferson:2021bid}, as well as to probe the physics of possible strongly-coupled chiral degrees of freedom associated to non-minimal F-theory singularities pointed out in~\cite{Candelas:2000nc,Hayashi:2008ba,Achmed-Zade:2018idx,Oehlmann:2021man,Jefferson:2021bid}.

Thanks to K\"ahler moduli independence, the Gukov--Vafa--Witten (GVW) superpotential~\cite{Gukov:1999ya} can also be computed readily from the geometry of the resolved elliptic CY background. The GVW superpotential takes the form
    \begin{equation}
        W = \int_{X^{(4)}} G_4 \wedge \Omega\,,
    \end{equation}
where $\Omega$ is the holomorphic 4-form of $X^{(4)}$, and can be written in terms of the periods of $\Omega$ in an integral basis of the horizontal 4-cycles. These periods are known to satisfy the Picard--Fuchs equations, and hence can be computed directly by solving these equations. However, directly computing the periods in an integral basis is rather technically involved. A much simpler computational strategy is to use mirror symmetry to recover the periods of $\Omega$ from the intersection numbers of a mirror CY 4-fold in the large volume/complex structure limit, see, e.g., \cite{Cota:2017aal,Marchesano:2021gyv}. Recall that mirror symmetry posits that certain pairs of type II string theories respectively describing strings propagating on so-called ``mirror pairs'' of CY manifolds $X^{(n)}, X^{(n)*}$ are dual to one another, satisfying
    \begin{equation}
        h^{p, q}(X^{(n)}) = h^{n - p, q}(X^{(n)*})\,.
    \end{equation}
A comprehensive review of this vast topic can be found in~\cite{MirrorSym}. The case of interest for the discussion at hand is the conjectured duality between type IIA string theory compactified on a smooth elliptic CY 4-fold $X^{(4)}$ and type IIA string theory compactified on the mirror CY 4-fold $X^{(4)*}$, which identifies the periods of $\Omega$ with the central charges of topological B-branes on $X^{(4)*}$.\footnote{In the context of type IIA compactified on an elliptic CY 4-fold, the GVW superpotential is identified with a 2D version of the 4D flux superpotential~\cite{Taylor:1999ii}.} With the aid of mirror symmetry, the task of computing the periods of $\Omega$ for an elliptic CY 4-fold $X^{(4)}$ in the large complex structure limit can be reduced to the task of computing quadruple intersection numbers of the mirror CY 4-fold $X^{(4)*}$ in the large volume limit.

\begin{center}
* \qquad\qquad * \qquad\qquad *
\end{center}

\noindent The above examples, while by no means exhaustive, nevertheless showcase a rich interplay between elliptic CY geometry and supersymmetric F-theory vacua. While this interplay is not unique to F-theory, the particular feature that makes F-theory special among various branches of string theory is the detailed correspondence between 7-brane gauge theory and local geometric singularities. This correspondence makes it possible to use gauge theory as an organizing principle for the geometric properties of elliptic CY $n$-folds. For example, the topological intersection numbers of an elliptic CY exhibit a tensor structure that depends on gauge-theoretic and representation-theoretic objects (see Appendix B of~\cite{Jefferson:2021bid} and references therein). The simplest example of such a tensor structure is the matrix of double intersections of Cartan divisors $\hat{D}_i$, $i = 1, \dotsc, \rank(G)$ of a smooth elliptic CY $n$-fold $X^{(n)} \to B^{(n - 1)}$ resolving a Kodaira singularity over $\Sigma \subset B^{(n - 1)}$:
    \begin{equation}
        \label{eq:gaugeexample}
        \hat{D}_i \hat{D}_j \hat{A} = -\kappa_{i j} \Sigma A\,,
    \end{equation}
where $\kappa_{i j}$ is the inverse metric tensor on $\Lie(G)$ and $A$ is the class of an $(n - 2)$-cycle in $B^{(n - 1)}$. From a ``low-energy'' perspective, gauge theory descriptions are highly redundant and non-unique, but in certain string vacua, gauge theory appears to enjoy a privileged role. F-theory, especially, seems to single out a particular gauge theory description of the effective $(12 - 2 n)$D theory describing the vacuum solution at low energies, which is necessarily independent of any choice of resolution of the singularities that preserves the CY condition. F-theory thus suggests that gauge theory provides a resolution-independent means of describing geometric and topological properties of singular CY varieties.\footnote{As \cref{eq:gaugeexample} demonstrates, a basic gauge-theoretic characterization of elliptic CY singularities that manifests itself in the tensor structure of intersection numbers is the characterization of codimension-one singular fibers (i.e., singular fibers over codimension-one components of the discriminant locus in $B^{(n - 1)}$) in terms of the inverse metric tensor of the gauge algebra. The inverse metric tensor is closely related to the Cartan matrix of the gauge algebra, and in fact there is a mathematical classification of codimension-one elliptic fiber singularities in terms of Cartan matrices due to Kodaira and Ner\'on; this classification is summarized in \cref{tab:Kodaira,tab:Tatetable}. Another example is the characterization of codimension-two elliptic fiber singularities in terms of representation-theoretic objects related to representations of charged matter in the low-energy spectrum transforming under $\G$; see, e.g., Eq.~(B.10) in~\cite{Jefferson:2021bid}, as well as~\cite{Hayashi:2014kca}.}

In light of the above discussion, one might wonder whether or not it is possible to compute geometric data like topological intersection numbers associated to singular elliptic CY $n$-folds without going through the trouble of resolving the singularities. While we do not attempt to address this question in this paper, we nevertheless point out that a very convenient way to bring the manifestly resolution-\emph{dependent} intersection numbers into contact with the resolution-invariant gauge theory data characterizing the F-theory vacuum is to use the projection formula (Eq.~(B.3) in~\cite{Jefferson:2021bid}) to ``integrate'' the intersection numbers over the elliptic fiber directions, which essentially entails computing the pushforward (defined with respect to the canonical projection $\pi\colon X^{(n)} \to B^{(n - 1)}$) of the intersection numbers to the Chow ring of the base:
    \begin{equation}
        \label{eq:pushintex}
        \pi_*\left(\int_{X^{(n)}} \hat{\omega}_{I_1} \wedge \hat{\omega}_{I_2} \wedge \dotsb \wedge \hat{\omega}_{I_n} \right) = W_{I_1 I_2 \dotsm I_n}^{\alpha_1 \alpha_2 \dotsm \alpha_{n - 1}} \int_{B^{(n - 1)}} \omega_{\alpha_1} \wedge \omega_{\alpha_2} \wedge \dotsb \wedge \omega_{\alpha_{n - 1}}\,.
    \end{equation}

Fortunately, it is indeed possible to evaluate the pushforward map $\pi_*$ for certain resolutions of singular elliptic CY varieties: namely, when $X^{(n)} \to X_0^{(n)}$ resolves a singular elliptic CY $X_0^{(n)}$ realized as a hypersurface of a projective bundle over a smooth projective $(n - 1)$-fold $B^{(n - 1)}$, and when the resolution $X^{(n)} \to X_0^{(n)}$ is comprised of a sequence of blowups along complete intersections of hyperplanes of the ambient space, then it is straightforward to evaluate $\pi_*$ explicitly using the methods of~\cite{Aluffi:2009tm,Fullwood:2011zb}, to which we refer the interested reader for comprehensive mathematical discussion and background. Quite satisfyingly, the resulting expression (i.e., the right-hand side of \cref{eq:pushintex}) only depends on the classes of certain divisors $L_a, \Sigma_s \in H_{1, 1}(B^{(n - 1)})$ defining the elliptic fibration and its discriminant locus, along with gauge-theoretic data characterizing the precise nature of the singular elliptic fibers. More recently, it was demonstrated~\cite{Esole:2017kyr} that the action of the pushforward map on any formal analytic function of intersection numbers can be evaluated via straightforward substitution, and hence any analytic function of the intersection numbers of an elliptic fibration can be converted into a formal analytic function of the classes $L_a, \Sigma_s$. This method for evaluating the pushforwards of intersection numbers has been used successfully to compute various topological invariants of families of elliptic $n$-folds called ``$\G$-models''~\cite{Esole:2018tuz,Esole:2018bmf}\footnote{The name ``$\G$-model'' is reference to the gauge group $\G$ associated to the codimension-one elliptic fiber singularities of the F-theory compactification.}, along with properties of numerous other supersymmetric vacua considered in various contexts, e.g.,~\cite{Bhardwaj:2018yhy,Bhardwaj:2018vuu,Jefferson:2021bid}.

Despite these successes, the applicability of the methods of~\cite{Esole:2017kyr} has so far been limited to elliptic CY $n$-folds corresponding to F-theory vacua with gauge groups $\G$ of relatively small rank, due to the computational expense involved in explicitly evaluating the pushforward map on individual intersection products as a means to compute topological intersection numbers. Since the singular fibers determine the gauge group, this technical obstruction has in turn limited the computation of topological intersection numbers via these methods to elliptic CY $n$-folds exhibiting a finite ``range'' of codimension-one elliptic fiber singularity types. The purpose of this paper is to provide an efficient tool for computing topological intersection numbers of elliptic CY $n$-folds with larger rank gauge groups, thus making topological properties of a wider range of singular elliptic CY $n$-folds accessible via the methods of~\cite{Esole:2017kyr}.

The remainder of this paper is structured as follows: in \cref{sec:main} we summarize the main results of the paper and provide a ``quick start'' for users who wish to skip the technical discussion and immediately get started using the \emph{Mathematica} package to perform computations. \Cref{sec:review} is a brief review of F-theory compactifications that provides context for the results of this paper. In \cref{sec:pushforwards}, we review how to compute the pushforwards of analytic functions of the classes of divisors in smooth or resolved singular elliptic fibrations to the base of the elliptic fibration, so that the analytic functions can be evaluated entirely in terms of the intersections of divisors in the base of the elliptic fibration. In \cref{sec:alg}, we describe the main result of this paper, which is the computational algorithm we use to evaluate the pushforward map explicitly. In \cref{sec:generating}, we describe how to use methods of \cref{sec:pushforwards} to compute the pushforward of the generating function of $n$-fold intersection numbers of divisors in resolved F-theory Tate models with gauge groups $\G = \SU(r + 1), \SO(2 r + 1), \Sp(r), \SO(2 r)$. We conclude in \cref{sec:discussion} with a discussion of various applications for the results of this paper along with possible future research directions. A compendium of notation used frequently throughout the paper is collected in \cref{sec:notation}.

\section{Summary and quick start guide}
\label{sec:main}

In this section, we summarize the main result and explain how to use the \emph{Mathematica} package \texttt{IntersectionNumbers.m} for computations.

\subsection{Overview}
In this paper, we present an efficient algorithm for computing topological intersection numbers of divisors in resolved CY $n$-folds $X^{(n)}$ elliptically fibered over a smooth K\"ahler base $B^{(n - 1)}$. Let us denote the canonical projection map to the base by $\pi$,
    \begin{equation}
        \pi\colon X^{(n)} \to B^{(n - 1)}\,.
    \end{equation}
A key step in this algorithm is organizing intersection products of divisors, $\hat{D}_I \subset X^{(n)}$ into a generating function given by the exponential of the K\"ahler class:
    \begin{equation}
        e^{\hat{J}}\,, \quad \hat{J} = \phi^I \hat{D}_I\,.
    \end{equation}
For a choice of integer coefficients $\phi^I$, the above generating function can be interpreted as the Chern character of the line bundle $\sO(\hat{J})$; however, we leave the coefficients $\phi^I$ variable. We then use the algorithm described in \cref{sec:alg} to judiciously apply the methods of~\cite{Esole:2017kyr} to compute the pushforward of $e^{\hat{J}}$ to the Chow ring of $B^{(n - 1)}$:
    \begin{equation}
        \label{eq:intropush}
        \pi_*(e^{\hat{J}}) = Z_{\phi}\,.
    \end{equation}
The pushforward replaces products of divisors $\hat{D}_I$ in $X^{(n)}$ with products of divisors $D_\alpha$ in $B^{(n - 1)}$; this is essentially ``integrating'' over the elliptic fiber directions:
    \begin{equation}
        \label{eq:pushonprod}
        \pi_*\left(\hat{D}_{I_1} \hat{D}_{I_2} \dotsm \hat{D}_{I_n}\right)
        = W_{I_1 I_2 \dotsm I_n}^{\alpha_1 \alpha_2 \dotsm \alpha_{n - 1} } D_{\alpha_1} D_{\alpha_2} \dotsm D_{\alpha_{n - 1}} \equiv W_{I_1 I_2 \dotsm I_n}\,,
    \end{equation}
where $W_{I_1 I_2 \dotsm I_n}^{\alpha_1 \alpha_2 \dotsm \alpha_{n-1}}$ are integer coefficients and $W_{I_1 I_2 \dotsm I_n}$ is a degree $n - 1$ homogeneous polynomial in certain divisor classes in the Chow ring of $B^{(n - 1)}$. This means that $n$-point intersection numbers of the elliptic CY $n$-fold $X^{(n)}$ can be evaluated in terms of $(n - 1)$-point intersection numbers of $B^{(n - 1)}$, again by taking derivatives:
    \begin{equation}
        \label{eq:derivative}
        W_{I_1 I_2 \dotsm I_k} = \left.\frac{\partial^k Z_\phi}{\partial\phi^{I_1}  \partial\phi^{I_2} \dotsm \partial\phi^{I_{k - 1}} \partial\phi^{I_k}}\right|_{\phi^I = 0}\,.
    \end{equation}

We implement our algorithm by means of a \emph{Mathematica} package \texttt{IntersectionNumbers.m} that computes the intersection numbers of resolutions
    \begin{equation}
        X^{(n)} \equiv X_r^{(n)} \to X_{r - 1}^{(n)} \to \dotsm \to X_1^{(n)} \to X_0^{(n)}
    \end{equation}
of singular elliptic CY $n$-folds $X_0^{(n)} \to B^{(n - 1)}$ that define F-theory Tate models (see~\cref{eq:Tate}) with gauge group
    \begin{equation}
        \G = \prod_s G_s\,, \quad r = \rank(\G)\,,
    \end{equation}
where $G_s$ are simple Lie groups, and we associate each simple group $G_s$ to a Kodaira singularity over a collection of codimension-one components of the discriminant locus,
    \begin{equation}
        \bigcup_s \Sigma_s \subset B^{(n - 1)}\,.
    \end{equation}
We assume for simplicity that the singular elliptic CY $n$-fold is constructed as a hypersurface of an ambient projective $(n + 1)$-fold
    \begin{equation}
        \label{eq:firstambient}
        Y_0^{(n + 1)} =\bP(\sL^{2} \oplus \sL^{3} \oplus \sO) \to B^{(n - 1)}\,, \quad \sL = \sK^{-1}\,,
    \end{equation}
where $\sK \to B^{(n - 1)}$ is the canonical line bundle over the base, and that the singularities have been resolved by means of a sequence of blowups along complete intersections of hyperplanes in $Y_0^{(n + 1)}$; note that to each of these blowups we associate the blowdown map
    \begin{equation}
        \label{eq:assocblowdown}
        f_i\colon Y_i^{(n + 1)} \to Y_{i - 1}^{(n + 1)}\,,
    \end{equation}
which contracts the exceptional divisor of the blowup in the ambient space.

Because $X^{(n)}$ has the structure of a fibration over a base $B^{(n - 1)}$, we can separate the divisor classes of $X^{(n)}$ into the classes of divisors that are pullbacks of divisors $D_\alpha \subset B^{(n - 1)}$, and classes $\hat{D}_{\hat{I}}$ that cannot be written as pullbacks of any divisor in the Chow ring of $B^{(n - 1)}$:
    \begin{equation}
        \hat{D}_I = \hat{D}_\alpha, \hat{D}_{\hat{I}}\,, \quad \hat{D}_\alpha \equiv \pi^*(D_\alpha)\,.
    \end{equation}
The pushforward $\pi_*$ only acts non-trivially on $\hat{D}_{\hat{I}}$; when acting on products that are comprised strictly of the classes $\hat{D}_{\hat{I}}$, $\pi_*$ produces a homogeneous polynomial that only depends on the canonical class $K = c_1(\sK)$ and the ``gauge divisors'' $\Sigma_s$ (compare to \cref{eq:pushonprod}):
    \begin{equation}
        \pi_*(\hat{D}_{\hat{I}_1} \hat{D}_{\hat{I}_2} \dotsm \hat{D}_{\hat{I}_n}) = \cW_{\hat{I}_1 \hat{I}_2 \dotsm \hat{I}_n}(K, \Sigma_s)\,.
    \end{equation}
Consequently, we may write $\hat{J} = \phi^\alpha \hat{D}_\alpha + \phi^{\hat{I}} \hat{D}_{\hat{I}}$, so that the pushforward of the generating function $e^{\hat{J}}$ takes the form
    \begin{equation}
        Z_\phi = \pi_*(e^{\hat{J}}) = e^{\phi^\alpha \hat{D}_\alpha} \cZ_\phi(K, \Sigma_s)\,, \quad \cZ_\phi(K, \Sigma_s) \equiv \pi_*(e^{\phi^{\hat{I}} \hat{D}_{\hat{I}}})\,,
    \end{equation}
where we emphasize that the analytic function $\cZ_\phi$ only depends on the divisor classes $K, \Sigma_s$, along with some parameters $\phi^{\hat{I}}$.

Given the above assumptions, the \emph{Mathematica} package \texttt{IntersectionNumbers.m} provides the function \texttt{push[generators, hypersurface]}, which accepts as its two inputs
    \begin{enumerate}
        \item \texttt{generators}: a two-dimensional array, consisting of the divisor classes of the hyperplanes whose complete intersections are blown up in the ambient spaces $Y_i^{(n + 1)}$ and;
        \item \texttt{hypersurface}: the divisor class of the resolved CY $n$-fold hypersurface $X^{(n)} \subset Y^{(n + 1)}$;
    \end{enumerate}
and returns the pushforward of the generating function $e^{\hat{J}}$. The divisor classes comprising the above input should be presented as linear combinations of the divisor classes $\bm{D}_\alpha, \bm{H}, \bm{E}_i$, where $\bm{D}_\alpha$ are the pullbacks of divisor classes in the Chow ring of $B^{(n - 1)}$, $\bm{H}$ is the hyperplane class of the $\bP^2$ fibers of the ambient space $Y_0^{(n + 1)}$ in \cref{eq:firstambient}, and $\bm{E}_i$ are total transforms of exceptional divisors that are contracted by the blowdown maps in \cref{eq:assocblowdown}. The latter two types of divisor classes can be represented in the \emph{Mathematica} package by the symbols \texttt{e[0]}, \texttt{e[1]}, \texttt{e[2]}, \textellipsis, \texttt{e[r]}, where \texttt{e[0]} is shorthand for the class $\bm{H}$. The symbols representing the classes $\bm{D}_\alpha$ can be arbitrary, as the pushforward map acts trivially on these classes.

\begin{figure}[!th]
    \centering

    \includegraphics[width=\textwidth]{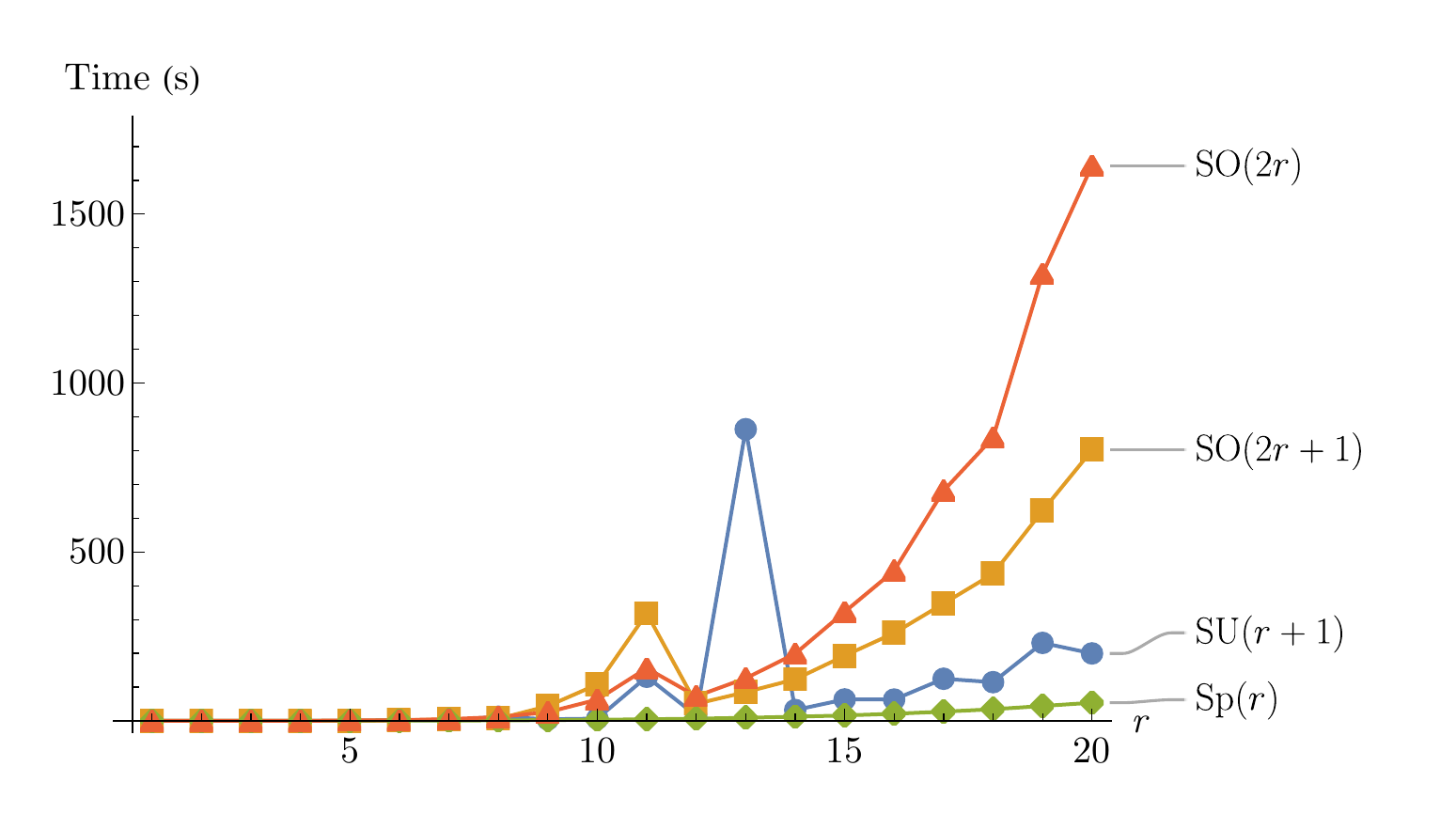}

    \caption{The above graph displays the time required to compute the pushforward of the generating function of intersection products of divisors (see \cref{eq:intropush}) of resolved F-theory Tate models by evaluating the function \texttt{push[]} in the \emph{Mathematica} package \texttt{IntersectionNumbers.m} with an 8-core Apple M1 CPU. The computation time is plotted as a function of the rank $r = \rank(\G)$ of the gauge group $\G = \SU(r + 1), \SO(2 r + 1), \Sp(r), \SO(2 r)$. Explicit expressions for these generating functions can be found in an ancillary \emph{Mathematica} notebook uploaded to the arXiv along with this preprint, and a few simple examples are also discussed in \cref{eq:Tate}. The sporadic jumps in computation time between $r - 1$ and $r + 1$ for certain values of $r$ (e.g., the $\SU(14)$ model) are due to the appearance of spurious poles that need to be canceled algebraically in order to carry out the symbolic substitutions required to evaluate \cref{eq:smoothpush,eq:singpush}---see \cref{sec:alg} for more details about the computational algorithm.}
    \label{fig:timegraph}
\end{figure}

We find that the algorithm upon which the function \texttt{push[]} is based significantly reduces the time necessary to carry out pushforward computations using the methods of~\cite{Esole:2017kyr}, as compared to available methods in the high energy theory literature. (As a qualitative example, the first author's personal machine, which uses a 2.6 GHz 6-Core Intel Core i7 processor, requires several hours to compute the quadruple intersection numbers for the $\SU(8)$ F-theory Tate model defined over a smooth 3-fold base using prior methods, whereas the methods of this paper can produce the same set of intersection numbers in under a minute.) This makes it possible to compute the intersection numbers of elliptic CY $n$-folds with large rank gauge groups.\footnote{Since the rank $r$ of the gauge group $\G$ is related to $h^{1, 1}(X^{(n)})$ by the Shioda--Tate--Wazir formula $h^{1, 1}(X^{(n)}) = 1 + h^{1, 1}(B^{(n - 1)}) + r$, it follows that our algorithm also in principle provides an efficient means to compute intersection numbers of elliptic CY $n$-folds with large $h^{1, 1}$.}
For a specific machine with given technical specifications, the time required to use the function \texttt{push[]} to evaluate the intersection numbers for all F-theory Tate models with gauge group $\G$ equal to one of the four simple classical Lie groups $\SU(r + 1), \SO(2 r + 1), \Sp(r), \SO(2 r)$ with rank $r \le 20$ is displayed in \cref{fig:timegraph}.

\subsection{Example: $\SU(2)$ model}

Before we begin, we must install the package and then load it into a \emph{Mathematica} notebook. The package can be installed using the \texttt{File > Install...} dialog (this only needs to be done once). The package should then be loaded into the current notebook using
    \begin{mmaCell}[moredefined=IntersectionNumbers,mathreplacements=light]{Code}
        <<IntersectionNumbers`;
    \end{mmaCell}
We are then ready to begin using the package.

As a simple example, we can consider the Tate-tuned $\SU(2)$ F-theory model, which is characterized by an $\singtype{I}_2$ singularity over the divisor $\Sigma \subset B^{(n - 1)}$ with local equation $\sigma = 0$; see \cref{sec:SUr+1model} for more details. This model can be resolved by a single blowup along the complete intersection of hyperplanes $x = y = \sigma = 0$; note that this blowup is one of a collection of standard blowup sequences for the simple classical Lie groups described in \cref{eq:Tate}, which is pre-loaded into the package \texttt{IntersectionNumbers.m}.

Note that the resolution $X^{(n)}$ has a basis of divisor divisors $\hat{D}_{\hat{I}} = \hat{D}_0, \hat{D}_1$, where
    \begin{equation}
        \hat{D}_0 = \bm{H} \cap \bm{X}^{(n)} / 3\,, \quad \hat{D}_1 = \bm{E}_1 \cap \bm{X}^{(n)}\,.
    \end{equation}
We would like to use the function $\texttt{push[]}$ to compute the pushforward of the generating function
    \begin{equation}
        e^{\hat{J}} = e^{\phi^\alpha \hat{D}_\alpha} e^{\phi^{\hat{I}} \hat{D}_{\hat{I}}} = e^{\phi^\alpha \bm{D}_\alpha} e^{\frac{1}{3} \alpha^0 \bm{H} + \alpha^1 \bm{E}_1} \cap \bm{X}^{(n)}\,, \quad \bm{X}^{(n)} = 3 \bm{H} - 6 \bm{K} - 2 \bm{E}_1\,.
    \end{equation}
Notice that we are working in the so-called $\alpha^I$ basis as opposed to the $\phi^I$ basis; in this case, these two bases are related by the map $\alpha^0 = \phi^0, \alpha^1 = \phi^1$. This change of basis can be computed using the function \texttt{aToPhi[\{algebra, rank\}..]}, where \texttt{algebra} is one of the four classical Lie algebra types (represented by the symbols \texttt{a}, \texttt{b}, \texttt{c}, \texttt{d}) and \texttt{rank} is the rank of the Lie algebra:\footnote{Although the parameters $\alpha^I, \phi^I$ in the \emph{Mathematica} package \texttt{IntersectionNumbers.m} appear with subscripts instead of superscripts, there is no distinction.}
    \begin{mmaCell}[moredefined=aToPhi]{Code}
        aToPhi[{a,1}]
    \end{mmaCell}
    \begin{mmaCell}{Output}
       \{\mmaSub{\(\alpha\)}{0}\;\(\rightarrow\)\;\mmaSub{\(\phi\)}{0},\;\mmaSub{\(\alpha\)}{1}\;\(\rightarrow\)\;\mmaSub{\(\phi\)}{1}\}
    \end{mmaCell}
The two inputs we require to compute the pushforward of $e^{\hat{J}}$ using the function \texttt{push[]} are the generators of the classes of the hyperplanes whose complete intersection is blown up, and the class of the hypersurface, $\bm{X}^{(n)}$. The former can be computed using the function \texttt{generators[]}:
    \begin{mmaCell}[moredefined=generators]{Code}
        generators[{a,1}]
    \end{mmaCell}
    \begin{mmaCell}{Output}
        \{\{2\;L\;+\;e[0],\;3\;K\;+\;e[0],\;s[1]\}\}
    \end{mmaCell}
The hypersurface class is computed using the function \texttt{hypersurface[]}:
    \begin{mmaCell}[moredefined=hypersurface]{Code}
        hypersurface[{a,1}]
    \end{mmaCell}
    \begin{mmaCell}{Output}
        6\;L\;+\;3\;e[0]\;-\;2\;e[1]
    \end{mmaCell}
The above input data are all that is needed to compute the generating function $e^{\phi^i \hat{D}_i}$ by evaluating
    \begin{mmaCell}[moredefined={push,generators,hypersurface}]{Code}
        push[generators[{a,1}],hypersurface[{a,1}]]
    \end{mmaCell}
    \begin{mmaCell}{Output}
        -\mmaFrac{\mmaSup{𝕖}{-L\;\mmaSub{\(\alpha\)}{0}}}{L}\;+\;\mmaFrac{-\mmaFrac{12\;\mmaSup{𝕖}{s[1]\;\mmaSub{\(\alpha\)}{1}}\;\mmaSup{L}{2}}{-2\;L\;+\;s[1]}\;+\;\mmaFrac{6\;\mmaSup{𝕖}{2\;L\;\mmaSub{\(\alpha\)}{1}}\;L\;s[1]}{-2\;L\;+\;s[1]}}{6\;\mmaSup{L}{2}}
    \end{mmaCell}
or alternatively, using the fact that the standard resolutions \labelcref{eq:Ares,eq:Bres,eq:Cres,eq:Dres} are built into the function \texttt{push[]}, one can simply enter
    \begin{mmaCell}[moredefined=push]{Code}
        push[{a,1}]
    \end{mmaCell}
    \begin{mmaCell}{Output}
        -\mmaFrac{\mmaSup{𝕖}{-L\;\mmaSub{\(\alpha\)}{0}}}{L}\;+\;\mmaFrac{-\mmaFrac{12\;\mmaSup{𝕖}{s[1]\;\mmaSub{\(\alpha\)}{1}}\;\mmaSup{L}{2}}{-2\;L\;+\;s[1]}\;+\;\mmaFrac{6\;\mmaSup{𝕖}{2\;L\;\mmaSub{\(\alpha\)}{1}}\;L\;s[1]}{-2\;L\;+\;s[1]}}{6\;\mmaSup{L}{2}}
    \end{mmaCell}
Note that when using the function \texttt{push[]}, the prefactor of $e^{\phi^\alpha \bm{D}_\alpha}$ is automatically omitted from the expression, since the pushforward acts trivially on functions of $\bm{D}_\alpha$; one can easily compensate for this omission by restoring a factor of $e^{\phi^\alpha D_\alpha}$ at the end of the computation, as we do in \cref{eq:earlySU2example} below.

The following expression is the output of the function \verb|push[{a,1}]| after making the substitutions $L \to -K, s[1] \to \Sigma$ and using \texttt{aToPhi[\{a,1\}]} to change variables:
    \begin{equation}
        \label{eq:earlySU2example}
        \cZ_\phi= -  \frac{\frac{2 K e^{\Sigma \phi^1} + \Sigma e^{-2 K \phi ^1}}{2 K + \Sigma } - e^{K \phi^0}}{K}\,.
    \end{equation}
Restoring an overall factor of $e^{\phi^\alpha D_\alpha}$, we get the complete generating function:
    \begin{equation}
        Z_\phi =  -e^{\phi^\alpha D_\alpha} \frac{\frac{2 K e^{\Sigma \phi^1} + \Sigma e^{-2 K \phi ^1}}{2 K + \Sigma } - e^{K \phi^0}}{K}\,.
    \end{equation}
One can extract intersection numbers by taking derivatives of the above expression. For example, in the case of a CY 4-fold, we could compute the intersections
    \begin{equation}
    \begin{aligned}
        \hat{D}_1^4 &= \left.\left(\frac{\partial}{\partial \phi^1}\right)^4 Z_\phi \right|_{\phi^I = 0} = -2 \Sigma \left(4 K^2 - 2 K \Sigma + \Sigma^2\right)\,, \\
        \hat{D}_0 \hat{D}_\alpha \hat{D}_\beta \hat{D}_{\gamma} &= \left.\frac{\partial}{\partial \phi^0} \frac{\partial}{\partial \phi^\alpha} \frac{\partial}{\partial \phi^\beta} \frac{\partial}{\partial \phi^\gamma} Z_\phi \right|_{\phi^I = 0} = \left.\frac{\partial}{\partial \phi^\alpha} \frac{\partial}{\partial \phi^\beta} \frac{\partial}{\partial \phi^\gamma} e^{\phi^\mu D_\mu} \right|_{\phi^\nu = 0} = D_\alpha D_\beta D_\gamma\,.
    \end{aligned}
    \end{equation}

\subsection{Function reference}

The package \texttt{IntersectionNumbers.m} provides the functions \texttt{push[]}, \texttt{aToPhi[]}, \texttt{generators[]}, \texttt{hypersurface[]}, and \texttt{divisors[]}. All of these functions accept any number of \texttt{\{algebra, rank\}} pairs, allowing one to compute intersection numbers for Tate models of arbitrary products of the simple classical Lie groups resolved using the blowup sequences described in \cref{eq:Tate}, e.g., \texttt{push[\{a, 2\}, \{b, 3\}, \{c, 4\}]}. The \texttt{?} command can be used to learn more about what these functions do and the inputs they accept, e.g., \texttt{?push}.

\section{Review of F-theory compactifications}
\label{sec:review}
In this section we review the physics background and primary motivation for the results of this paper.

\subsection{Basic setup}
F-theory is an $\SL(2, \bZ)$ duality-symmetric geometric formulation of type IIB string theory compactified on a K\"ahler manifold $B^{(n - 1)}$ of complex dimension $n - 1$ in the presence of 7-branes and O7-planes~\cite{Vafa:1996xn,Morrison:1996na,Morrison:1996pp}. In this formulation, the vacuum profile of the axiodilaton field $\tau$ appearing in the $\SL(2, \bR)$-symmetric low-energy effective type IIB supergravity action is interpreted as the complex structure of a genus one complex curve (which in our case we take to be an elliptic curve---see \cref{foot:genusone}) fibered over the space $B^{(n - 1)}$, where the subloci in $B^{(n - 1)}$ over which the elliptic fiber degenerates are interpreted as the locations of 7-branes. Compatibility with supersymmetry implies that the total space of the singular elliptic fibration $X_0^{(n)} \to B^{(n - 1)}$ must be (the possibly singular limit of) a manifold with $\SU(n)$ holonomy, and hence we take $X_0^{(n)}$ to be a singular elliptically fibered CY $n$-fold. Formally viewed as a 12D theory, F-theory compactified on an elliptically fibered CY $n$-fold is therefore described at low energies by a supergravity theory on $\bR^{1, 11 - 2 n}$ preserving at least $2^{6 - n}$ supercharges; see \cref{tab:SUSY}.

\begin{table}[!th]
    \centering

    \begin{tabular}{|c|c|c|}\hline
        $n$ & SUSY algebra & \# of supercharges \\ \hline\hline
        1 & 10D  $\cN = 2$ & 32 \\ \hline
        2 & 8D $\cN = 1$ & 16 \\ \hline
        3 & 6D $\cN = (1, 0)$ & 8 \\ \hline
        4 & 4D $\cN = 1$ & 4 \\ \hline
        5 & 2D $\cN = (2, 0)$ & 2 \\ \hline
    \end{tabular}

    \caption{Table of lower-dimensional supersymmetry (SUSY) algebras preserved by F-theory compactified on a singular CY $n$-fold $X_0^{(n)}$ elliptically fibered over a smooth K\"ahler space $B^{(n - 1)}$ (note that we assume $X_0^{(n)}$ has $\SU(n)$ holonomy). The special cases $n = 1, 2$ correspond, respectively, to a torus $T^2$ and a K3 surface. Special choices of compactification geometry can preserve an enhanced supersymmetry algebra; for example, compactification on a smooth torus $T^{2 n}$, which can be interpreted as a trivial elliptic fibration over $B^{(n - 1)} = T^{2 n - 2}$, has trivial holonomy and hence preserves maximal supersymmetry. In the right-most column, we indicate the number of conserved supercharges generating the supersymmetry algebra. Cases for which $n > 5$ do not have a clear physical interpretation.}
    \label{tab:SUSY}
\end{table}

As noted in \cref{sec:intro}, singular elliptic CY varieties play a central role in constructing F-theory vacua, and the interplay between their mathematical properties and the physics of F-theory vacua is the principal motivation for the results of this paper. We next review some additional background for F-theory compactifications in order to provide context for the computations we describe in later sections.

\subsection{Singular elliptic fibrations and Weierstrass models}
The elliptic fibration $X_0^{(n)} \to B^{(n - 1)}$ defining an F-theory vacuum is conventionally described explicitly as a generically singular Weierstrass model
    \begin{equation}
        \label{eq:Weierstrass}
        y^2 - (x^3 + f x + g) = 0
    \end{equation}
along with a choice of characteristic line bundle
    \begin{equation}
        \label{eq:charbundle}
        \sL \to B^{(n - 1)}\,.
    \end{equation}
In \cref{eq:Weierstrass}, $x, y$ are complex variables and $f, g$ are sections of the line bundles (resp.) $\sL^{\otimes 4}, \sL^{\otimes 6}$. In order to impose the CY condition, namely the condition that the first Chern class of $X^{(n)}$ vanishes\footnote{Note that vanishing first Chern class (which is equivalent to Ricci-flatness) implies that $X^{(n)}$ admits a finite cover with trivial canonical bundle, and hence global $\SU(n)$ holonomy \cite{beauville1983varietes}.}, we take the characteristic line bundle in \cref{eq:charbundle} to be equal to the anticanonical bundle of $B^{(n - 1)}$,
    \begin{equation}
        \label{eq:CYcondition}
        \sL = \sK^{-1}\,.
    \end{equation}
The elliptic fibration described by \cref{eq:Weierstrass} is not in general smooth, and has singularities associated to the discriminant locus
    \begin{equation}
        \Delta \equiv 4 f^3 + 27 g^2 = 0
    \end{equation}
in $B^{(n - 1)}$. The kinematic structure of the low-energy gauge sector of the F-theory compactification is encoded in the singularities of the elliptic fiber over various components of the discriminant locus. For instance, the possible singularity types of the elliptic fiber over components of $\Delta = 0$ of maximal dimension in $B^{(n - 1)}$ (i.e., codimension one in $B^{(n - 1)}$) have been classified by Kodaira and N\'eron~\cite{KodairaII,KodairaIII,NeronClassification} (roughly) in terms of the simple Lie algebras, and determine the gauge algebra $\Lie(\G)$ of the low-energy effective theory.

Suppose that a codimension-one component of $\Delta = 0$ can be described, in local coordinates, by the equation $\sigma = 0$. Then, the singularity type of the elliptic fiber can be determined by expanding the sections $f, g, \Delta$ as power series in $\sigma$,
    \begin{equation}
        f = f_m \sigma^m\,, \quad g = g_m \sigma^m\,, \quad \Delta = \Delta_m \sigma^m\,,
    \end{equation}
and checking the orders of vanishing of $f, g, \Delta$ in a neighborhood of the locus $\sigma = 0$ (the orders of vanishing $\ord(f), \ord(g), \ord(\Delta)$ are given by the smallest powers of $\sigma$ for which $f_m, g_m, \Delta_m$ are respectively non-vanishing). This leads to a (nearly) one-to-one identification between the orders of vanishing $\ord(f), \ord(g), \ord(\Delta)$ characterizing the singularities of the elliptic fiber in codimension one, and simply laced Lie algebras---see \cref{tab:Kodaira}. It is possible to give a more refined classification of the codimension-one elliptic fiber singularities by specifying both the orders of vanishing of $f, g, \Delta$ and their (Tate) monodromy around the codimension-one components of $\Delta = 0$; in cases with non-trivial such monodromy, the singularity is instead associated with a non-simply laced Lie algebra given by a particular folding of the original simply laced algebra. For certain types of algebro-geometric constructions, these monodromy constraints can be straightforwardly imposed; a family of such constructions relevant to the results of this paper are the Tate models~\cite{Bershadsky:1996nh,Katz:2011qp}, which we describe in more detail in \cref{eq:Tate}. Additional kinematic properties of the low-energy theory are encoded in elliptic fiber singularities over higher-codimension subloci of $\Delta = 0$; see, e.g., Table~1.1 of~\cite{Weigand:2018rez} for a summary.

\begin{table}[!th]
    \centering

    $
    \begin{array}{|c|c|c|c|c|c|}\hline
        \text{fiber type} & \ord(f) &\ord(g) & \ord(\Delta) & \Lie(\G_\text{ADE}) & \text{monodromy} \\ \hline\hline
        \singtype{I}_0         & \ge 0 & \ge 0 & 0      & \text{trivial} & \text{trivial} \\ \hline
        \singtype{I}_{r + 1}   & 0     & 0     & r + 1  & \gA_r      & \begin{pmatrix} 1  & r + 1 \\ 0  & 1  \end{pmatrix} \\ \hline
        \singtype{II}          & \ge 1 & 1     & 2      & \text{---} & \begin{pmatrix} 1  & 1     \\ -1 & 0  \end{pmatrix} \\ \hline
        \singtype{III}         & 1     & \ge 2 & 3      & \gA_1      & \begin{pmatrix} 0  & 1     \\ -1 & 0  \end{pmatrix} \\ \hline
        \singtype{IV}          & \ge 2 & 2     & 4      & \gA_2      & \begin{pmatrix} 0  & 1     \\ -1 & -1 \end{pmatrix} \\ \hline
        \singtype{I}_0^*       & \ge 2 & \ge 3 & 6      & \gD_4      & \begin{pmatrix} -1 & 0     \\ 0  & -1 \end{pmatrix} \\ \hline
        \singtype{I}_{r - 4}^* & 2     & 3     & r + 2  & \gD_r      & \begin{pmatrix} -1 & 4 - r \\ 0  & -1 \end{pmatrix} \\ \hline
        \singtype{IV}^*        & \ge 3 & 4     & 8      & \gE_6      & \begin{pmatrix} -1 & -1    \\ 1  & 0  \end{pmatrix} \\ \hline
        \singtype{III}^*       & 3     & \ge 5 & 9      & \gE_7      & \begin{pmatrix} 0  & -1    \\ 1  & 0  \end{pmatrix} \\ \hline
        \singtype{II}^*        & \ge 4 & 5     & 10     & \gE_8      & \begin{pmatrix} 0  & -1    \\ 1  & 1  \end{pmatrix} \\ \hline
        \text{non-min}         & \ge 4 & \ge 6 & \ge 12 & \text{?}   & \text{?}                                            \\ \hline
    \end{array}
    $

    \caption{Possible elliptic fiber singularity types that can develop over irreducible codimension-one components of the discriminant locus $\Delta = 0$ in $B^{(n - 1)}$. The singularity types are specified in terms of the order of vanishing of the sections $f, g, \Delta$, expressed locally as polynomials in the complex coordinates of $B^{(n - 1)}$. The right-most column indicates the $\SL(2, \bZ)$ monodromy experienced by the axiodilaton $\tau$ encircling the codimension-one component of the discriminant locus, which signals the presence of a configuration of 7-branes and O7-planes. (This is not to be confused with the Tate monodromy of the resolved fiber components around the codimension-one components of the discriminant locus.) The final row describes ``non-minimal'' singularities in F-theory, which by themselves do not admit resolutions preserving the CY condition corresponding to flat elliptic fibrations; the physics of these singularities can in some circumstances be explored by studying K\"ahler or complex deformations of the base of the elliptic fibration~\cite{Morrison:2014lca}.}
    \label{tab:Kodaira}
\end{table}

A standard approach in algebraic geometry for analyzing the singularities of a projective variety is to (non-uniquely) resolve the singularities by means of a sequence of blowups, whereby the singular loci are replaced by smooth K\"ahler varieties. To preserve the physical interpretation of $X_0^{(n)}$ as a string theory vacuum solution, we must choose our sequence of blowups such that the first Chern class of $X_0^{(n)}$ remains unchanged, i.e., so that the CY condition is preserved. In favorable cases, a finite sequence of blowups with this property can be identified, leading to a resolution of singularities 
	\begin{align}
		X^{(n)} \to X_0^{(n)}.
	\end{align}
We thus turn our attention to analyzing the geometric and topological properties of smooth elliptic CY $n$-folds $X^{(n)}$ resolving a singular elliptic CY $n$-fold $X_0^{(n)}$.

As discussed in \cref{sec:intro}, a particularly useful set of topological invariants for analyzing the physics of the low-energy effective theory of an F-theory compactification are the $n$-fold intersection numbers of divisors $\hat{D}_I \in H_{n - 1,n - 1}(X^{(n)})$, where we view divisors as the Poincar\'e duals of harmonic forms $\omega_I \in H^{1, 1}(X^{(n)})$; see \cref{basicint}. The intersection numbers of divisors can be interpreted as counting, with signs and multiplicities, the number of points of intersection of oriented holomorphic $n - 1$ cycles in $X^{(n)}$. Computing the intersection numbers of elliptic fibrations is the main subject of this paper, and will be the focus of the remaining sections. 

\section{Pushforwards of analytic functions of intersection products}
\label{sec:pushforwards}

In this section, we take a step back from the specific setting of F-theory geometry and review a well-known strategy for computing topological intersection numbers in resolutions of certain singular varieties that carry the structure of a fibration, namely by computing the pushforward of the intersection numbers to the base of the fibration.

\subsection{Overview}
\label{sec:pushover}
Consider a smooth $n$-fold $X^{(n)}$ possibly resolving a singular $n$-fold $X_0^{(n)}$. Let $X^{(n)}$ have a basis of divisors $\hat{D}_{I}$, $I = 1, \dotsc, h^{1, 1}(X^{(n)})$, and assume that $X^{(n)}$ is equipped with a projection
    \begin{equation}
        \label{eq:canonicalprojection}
        \pi\colon X^{(n)} \to B^{(m)}\,, \quad m < n\,,
    \end{equation}
so that $X^{(n)}$ has the structure of a fibration.

In this section, we review how to evaluate an arbitrary formal analytic function $F$ in the Chow ring of $X^{(n)}$, i.e., a function of the form\footnote{The coefficients $F_{p_1, p_2, \dotsc, p_{h^{1, 1}}}$ are complex numbers, while the graded terms (i.e., terms of fixed degree $k = \sum p_I \le n$) they multiply in \cref{eq:analytic} correspond to the $k$-fold products of divisor classes in the Chow ring of $X^{(n)}$. Equivalently, given a class $A$ of degree $n - k = \sum q_I$ in the Chow ring of $X^{(n)}$, we may think of the degree $k$ terms in \cref{eq:analytic} as being defined by the intersection numbers
    \begin{equation}
        A \hat{D}_1^{p_1} \hat{D}_2^{p_2} \dotsm \hat{D}_{h^{1, 1}}^{p_{h^{1, 1}}} = \int_{X^{(n)}} \PD(A) \wedge \omega_1^{p_1} \wedge \omega_2^{p_2} \wedge \dotsb \wedge \omega_{h^{1, 1}}^{p_{h^{1, 1}}} \in \bZ\,,
    \end{equation}
where `$\PD$' denotes the Poincar\'e dual. Note that the subscripts $h^{1, 1}$ are shorthand for $h^{1, 1}(X^{(n)})$.}
    \begin{equation}
        \label{eq:analytic}
        F(\hat{D}_I) = \sum_{p_1, p_2, \dotsc, p_{h^{1, 1}}} F_{p_1, p_2, \dotsc, p_{h^{1, 1}}} \hat{D}_1^{p_1} \hat{D}_2^{p_2} \dotsm \hat{D}_{h^{1, 1}}^{p_{h^{1, 1}}}\,
    \end{equation}
where in the above equation the subscripts $h^{1, 1}$ are shorthand for $h^{1, 1}(X^{(n)})$. We may regard $F(\hat{D}_I)$ as a characteristic class, analogous to, e.g., the Chern polynomial. If we rescale all of the divisors, $\hat{D}_I \to \varepsilon \hat{D}_I$, we may write
    \begin{equation}
        \label{Fhat}
        F(\varepsilon \hat{D}_I) = \sum_k F_k(\hat{D}_I) \varepsilon^k\,,
    \end{equation}
so that the coefficient of $\varepsilon^k$ in the above power series expansion, $F_k(\hat{D}_I)$, is a homogeneous degree-$k$ polynomial in the classes $\hat{D}_I$.

Our strategy for evaluating the arbitrary function $F(\hat{D}_I)$ described above is to compute the pushforward of $F(\hat{D}_I)$ to $B^{(m)}$, so that the intersection products comprising the summands of the formal power series expansion in \cref{eq:analytic} can be expressed as intersection products of divisors $D_{\alpha}$ in the Chow ring of $B^{(m)}$, following \cref{eq:pushonprod}. Acting term-by-term in the above power series expansion, this produces a new power series:
    \begin{equation}
        \pi_*( F(\hat{D}_I)) \equiv \pi_*(F)(D_\alpha) = \sum_{q_1, q_2, \dotsc, q_{h^{1, 1}}} \pi_*(F)_{q_1, q_2, \dotsc, q_{h^{1, 1}}} D_1^{q_1} D_2^{q_2} \dotsm D_{h^{1, 1}}^{q_{h^{1, 1}}}\,
    \end{equation}
where we note that the subscripts $h^{1, 1}$ in the above equation are shorthand for $h^{1, 1}(B^{(m)})$. If we rescale all of the divisors, $D_\alpha \to \varepsilon D_\alpha$, we may write
    \begin{equation}
        \label{Fpush}
        \pi_*(F)(\varepsilon D_\alpha) = \sum_k \pi_*(F)_k(D_\alpha) \varepsilon^k\,,
    \end{equation}
where $\pi_*(F)_k(D_\alpha)$ is a homogeneous degree-$k$ polynomial in the classes $D_\alpha$. We can match the power series \cref{Fhat} and \cref{Fpush} term-by-term provided we carefully account for the difference in dimension of the two spaces $X^{(n)}$ and $B^{(m)}$.\footnote{Given a proper morphism of varieties $f\colon X \to X'$ and a subvariety $Z \subset X$, the pushforward $f_*(Z) = 0$ if $\dim(f(Z)) < \dim(Z)$; see~\cite{Esole:2017kyr} and references therein for more mathematical background.} Thus, a more refined description of the action of the pushforward map is
    \begin{equation}
        \pi_*(F_k(\hat{D}_I)) = \pi_*(F)_{k - n+m}(D_\alpha)\,.
    \end{equation}
In the special case $k = n$, we see that both sides of the above equation correspond to well-defined intersection numbers, with the argument of left-hand side corresponding to an intersection product in the Chow ring of $X^{(n)}$ and the right-hand side corresponding to an intersection product in the Chow ring of $B^{(m)}$. We assume that the intersection numbers of $B^{(m)}$ are known, so that elements of fixed degree appearing on the right-hand side of the above equation can be readily computed.

In this paper, we focus on the special case that $X^{(n)}$ is an elliptic fibration, so that $m = n - 1$, i.e., $B^{(m)} = B^{(n - 1)}$ has dimension one less than the dimension of $X^{(n)}$. We assume that $X^{(n)}$ can be constructed as a hypersurface of a smooth ambient projective variety $Y^{(n + 1)}$. We furthermore assume that anytime $X^{(n)}$ resolves a singular elliptic fibration $X^{(n)}_0\subset Y_0^{(n + 1)}$, the resolution $X^{(n)} \to X_0^{(n)}$ can be decomposed into a sequence of blowups of another smooth ambient projective variety $Y_0^{(n + 1)}$ along complete intersections of hyperplanes, where we denote by $f_i$ the blowdown map
    \begin{equation}
        \label{eq:ambientblowdowns}
        f_i\colon Y_i^{(n + 1)} \to Y_{i - 1}^{(n + 1)}
    \end{equation}
that contracts the exceptional divisor $e_i = 0$ of the blowup of $Y_{i - 1}^{(n + 1)}$. The singular variety $X_0^{(n)}$ is taken to be a hypersurface of $Y_0^{(n + 1)}$, which is equipped with a canonical projection
    \begin{equation}
        \label{eq:canpro}
        \varpi\colon Y_0^{(n + 1)} \to B^{(n - 1)}\,.
    \end{equation}
In particular, we take $Y^{(n + 1)}_0$ to be the projectivization of a rank two vector bundle,
    \begin{equation}
        Y^{(n + 1)}_0 = \bP(\sV)\,, \quad \sV = \bigoplus_{a = 1}^3 \sL_a\,.
    \end{equation}
When $X^{(n)}$ is a smooth elliptic fibration, we simply write $X^{(n)} = X^{(n)}_0$ (and likewise $Y^{(n+1)} = Y_0^{(n+1)}$). Finally, we assume that the Chow ring classes of all divisors $\hat{D}_I \subset X^{(n)}$ can be expressed as the restrictions of divisor classes in the Chow ring of $Y^{(n + 1)}$ to the class of the hypersurface $X^{(n)} \subset Y^{(n + 1)}$. This ensures that intersection products in the Chow ring of $X^{(n)}$ can be expressed in terms of intersection products in the Chow ring of $Y^{(n + 1)}$.

Given the above assumptions, the central formula of this section is an explicit expression for the lift of the pushforward map $\pi_*$ to the ambient space $Y^{(n + 1)}$ of the resolution $X^{(n)}$. The lift of this map to the ambient space can be expressed in terms of a composition of pushforward maps $f_{i *}, \varpi_*$, where $f_i$ are the blowdown maps \labelcref{eq:ambientblowdowns} and $\varpi$ is the canonical projection \labelcref{eq:canpro}:
    \begin{equation}
        \label{eq:comp}
        \varpi_* \circ f_{1 *} \circ \dotsb \circ f_{r *}\,.
    \end{equation}

The fact that $X^{(n)}$ has the structure of a fibration means that we can separate the classes of divisors in $X^{(n)}$ into those that are the classes of pullbacks of divisors in the base, and those that are not,
    \begin{equation}
        \hat{D}_I = \hat{D}_\alpha, \hat{D}_{\hat{I}}\,, \quad \hat{D}_\alpha \equiv \pi^*(D_\alpha)\,.
    \end{equation}
This distinction is useful because the pushforward map $\pi_*$ only acts non-trivially on divisors $\hat{D}_{\hat{I}}$, i.e., $\pi_*(\hat{D}_\alpha) = \pi_*(\pi^*(D_\alpha)) = D_\alpha$. Thus, for the purposes of evaluating the action of $\pi_*$, we treat the pullbacks $\hat{D}_\alpha$ as ``spectators'' akin to numerical coefficients. For simplicity of exposition, unless otherwise specified, we only consider functions $F(\hat{D}_{\hat{I}})$ of the classes $\hat{D}_{\hat{I}}$ upon which $\pi_*$ acts non-trivially.\footnote{However, when computing pushforwards defined with respect to individual maps in the composition \labelcref{eq:comp}, we often abuse notation and write the representative of the function in the Chow ring of the ambient space $Y_i^{(n + 1)}$ as only depending on the class that is acted upon non-trivially by the pushforward map, despite the fact that the function may have non-trivial dependence on other divisor classes---see \cref{eq:analyticlift} for an example of this abuse of notation.} This ensures that the resulting pushforward is an analytic function that only depends on the classes of special divisors in the base related to the hypersurface embedding of $X_0^{(n)}$ in $Y_0^{(n + 1)}$. In particular, we find that the pushforward of $F(\hat D_{\hat I})$ only depends on $L_a = c_1(\sL_a)$ and the divisor classes of certain irreducible components $\Sigma_s$ of the discriminant locus of $B^{(n - 1)}$ that dictate the structure of the singularities of $X_0^{(n)}$:
    \begin{equation}
        \pi_*(F(\hat{D}_{\hat{I}})) = \cF(L_a, \Sigma_s)\,.
    \end{equation}

 In \cref{sec:smooth}, we review how to compute the pushforward with respect to the map $\varpi$, which is the only relevant map in the case of a smooth elliptic fibration; in \cref{sec:pushsing}, we review the generalization of this procedure to resolutions of singular elliptic fibrations, in which case we require the pushforwards defined with respect to the blowdown maps $f_i$. Note that \cref{sec:smooth,sec:pushsing} are a summary of the results of~\cite{Esole:2017kyr}, to which we refer the interested reader for detailed mathematical background and references. Later, in \cref{sec:alg}, we describe the main result of this paper, which is an efficient computational algorithm that can be used to evaluate the pushforward maps $\varpi_*, f_{i *}$ with the aid of a symbolic computing tool.

\subsection{Complete intersections of a smooth projective bundle}
\label{sec:smooth}

We first consider the case of a smooth $n$-fold $X_0^{(n)}$, and for now we relax the assumption that $X_0^{(n)}$ is an elliptic fibration. Our strategy for evaluating analytic functions like $F(\hat{D}_{\hat{I}})$ described in \cref{eq:analytic} is to compute the pushforward of $F(\hat{D}_{\hat{I}})$ to the Chow ring of the base $B^{(n - 1)}$, producing another analytic function that only depends on a special subset of divisor classes $L_a$ in $B^{(n - 1)}$:
    \begin{equation}
        \label{eq:genpush}
        \pi_*(F(\hat{D}_{\hat{I}})) = \cF(L_a)\,.
    \end{equation}
For simplicity, we restrict our discussion to $n$-folds that can be constructed explicitly as hypersurfaces $X_0^{(n)} \subset Y_0^{(n + 1)}$ of rank-two projective bundles of the form
    \begin{equation}
        Y_0^{(n + 1)} = \bP(\sV)\,, \quad \sV = \sL_1 \oplus \sL_2 \oplus \sL_3
    \end{equation}
equipped with the canonical projection
    \begin{equation}
        \varpi\colon Y_0^{(n + 1)} \to B^{(n - 1)}\,.
    \end{equation}
The generalization to complete intersections of smooth ambient projective bundles of arbitrary rank is straightforward to compute, albeit more tedious. 

The divisor classes
    \begin{equation}
        L_a = c_1(\sL_a) = L_a^\alpha D_\alpha
    \end{equation}
can be regarded as characteristic data defining the projective bundle $Y_0^{(n + 1)}$. Note that the ambient projective bundle $Y_0^{(n + 1)}$ has a natural basis of divisors
    \begin{equation}
        \bm{D}_\alpha, \bm{H}\,,
    \end{equation}
where $\bm{D}_\alpha \equiv \varpi^* D_\alpha$ are the pullbacks of divisors $D_\alpha \in H_{1, 1}(B^{(n - 1)})$, and $\bm{H}$ is the hyperplane class\footnote{That is, we introduce homogeneous coordinates $ [y_1 : y_2 : y_3]$ for the fibers of $Y_0^{(n + 1)}$ such that $\bm{y}_a = y_a^0 \bm{H} + y_a^\alpha \bm{D}_\alpha$, where $\bm{H} \equiv c_1(\sO_{\bP(\sV)}(1))$ is the class of the divisor associated to Serre's twisting sheaf $\sO_{\bP(\sV)}(1)$. These divisor classes can be viewed as hyperplane classes of the $\bP^2$ fibers twisted by pullbacks of the classes of certain divisors in $B^{(n - 1)}$ as necessary to construct $X^{(n)}$ as a hypersurface of $Y^{(n + 1)}$. \label{foot:ydiv}} of the $\bP^2$ fibers of $Y_0^{(n + 1)}$.

The pushforward $\pi_*$ can be lifted to a pushforward defined with respect to the ambient space projection map $\varpi$ as follows: in well-behaved constructions, the classes of divisors $\hat{D}_{\hat{I}} \in X_0^{(n)}$ can written as the restriction of divisors in $Y_0^{(n + 1)}$ to the class of the hypersurface $X_0^{(n)}$:
    \begin{equation}
        \label{eq:hyprestrict}
      \hat{D}_{\hat{I}} = \hat{\bm{D}}_{\hat{I}} \cap \bm{X}^{(n)}
    \end{equation}
where we regard the classes $\hat{\bm{D}}_{\hat{I}}, \bm{X}^{(n)}$ as linear combinations of the classes $\bm{L}_a = L_a^\alpha \bm{D}_\alpha$ and $\bm{H}$ of $Y_0^{(n + 1)}$ that depend on the details of the embedding of $X^{(n)}$ as a hypersurface in $Y_0^{(n + 1)}$. The above definition extends to arbitrary products of divisors $\hat{D}_{\hat{I}}$, and by linearity, also extends to formal infinite series~\cite{Esole:2017kyr}. Thus, any analytic function of the divisor classes $\hat{D}_{\hat{I}}$ can be written as
    \begin{equation}
        \label{eq:analyticlift}
        F(\hat{D}_{\hat{I}}) = F(\hat{\bm{D}}_{\hat{I}}) \cap \bm{X}^{(n)} \equiv \tilde{F}(\bm{H})\,.
    \end{equation}
Notice that on the right-hand side of the above equation, we are abusing notation and writing $\tilde{F}$ as a function only of the class $\bm{H}$, even though $\tilde{F}$ also can in principle depend on the pullbacks $\bm{L}_a$. We then use the fact that the action of the pushforward map $\varpi_*$ on an arbitrary analytic function of the hyperplane class $\bm{H}$ is known (see Appendix~E.1 in~\cite{Jefferson:2021bid})\footnote{We assume here generically $L_a \ne L_b$ for $a \ne b$. When $L_a = L_b$, we evaluate the above expression as a limit to ensure that spurious poles cancel.}:
    \begin{equation}
        \label{eq:smoothpush}
        \pi_*(F(\hat{D}_{\hat{I}})) = \varpi_*(\tilde{F}(\bm{H})) = \sum_{a = 1}^3 \frac{\tilde{F}(-L_a)}{\prod_{b \ne a} (L_a - L_b)} \equiv \cF(L_a)\,.
    \end{equation}
As discussed in \cref{sec:pushover}, we treat $\tilde{F}$ as an analytic function of $\bm{H}$ since the projection formula implies that the pushforward map $\varpi_*$ only acts non-trivially on the class $\bm{H}$.

\subsubsection{Example: projective bundle over projective space}

We illustrate the above formula using the simplest example of a hypersurface of a rank-two projective bundle, namely a smooth degree-$d$ hypersurface of a $\bP^2$ bundle over $\bP^{n - 1}$,
    \begin{equation}
        X_0^{(n)} \subset Y_0^{(n + 1)} = \bP(\sV)\,, \quad \sV = \sO(s) \oplus \sO(t) \oplus \sO\,.
    \end{equation}
Here, we make the identifications
    \begin{equation}
        \bm{H} = c_1(\sO_{\bP(\sV)}(1))\,, \quad L_1 = s J\,, \quad L_2 = t J\,,
    \end{equation}
where $J$ is the hyperplane class of $\bP^{n - 1}$, satisfying
    \begin{equation}
        J^{n - 1} = 1\,.
    \end{equation}
Thus the ambient space $Y_0^{(n + 1)}$ has a basis of divisors $\bm{H}, \bm{J}$, and the hypersurface $X_0^{(n)}$ has a basis of divisors $\hat{H} = \bm{H} \cap \bm{X}_0^{(n)}, \hat{J} = \bm{J} \cap \bm{X}_0^{(n)}$ where $\bm{X}_0^{(n)} = d \bm{H}$. The relevant pushforwards are:
    \begin{equation}
        \varpi_*(\bm{H}^p) = \frac{(-s J)^p}{(s J - t J)(s J)} + \frac{(-t J)^p}{(t J - s J)(t J)} = S_{p - 2}(-s J, -t J)\,, \quad p \ge 1\,,
    \end{equation}
where $S_q$ is the homogeneous symmetric polynomial of degree $q$ and we adopt the convention $S_{-1} = 0$. Hence, we can write any intersection number in $X_0^{(n)}$ as
    \begin{equation}
        \pi_*(\hat H^{q} \hat{J}^{n - q}) = \varpi_*((\bm{H})^q (\bm{J})^{n - q} (d \bm{H})) = d S_{q - 1}(-s J, -t J) J^{n - q}\,.
    \end{equation}
The above formulas can be extended to formal analytic functions in a straightforward fashion.

\subsection{Singular hypersurfaces}
\label{sec:pushsing}

We next consider cases in which $X_0^{(n)}$ is a singular variety, and we assume it is possible to obtain a smooth $n$-fold $X^{(n)}$ from $X_0^{(n)}$ by way of a finite sequence of blowups.\footnote{For string compactifications, we both need to impose the CY condition on the singular variety and ensure that its blowups preserve the CY condition. The first Chern class of a smooth projective variety $X$ (which is also a K\"ahler manifold) is equal to the first Chern class of the anticanonical line bundle. For some types of singular varieties $X_0$ (e.g., hypersurfaces of smooth varieties), the canonical class $K_{X_0}$ is still well-defined, and one can impose the CY condition by demanding that $K_{X_0} = 0$. Preserving the CY condition in these cases means resolving the singularities of $X_0$ by means of sequences of blowups that do not change $K_{X_0}$, leading eventually to a resolution $X \to X_0$ with vanishing first Chern class, $c_1(X) = -K_X = - K_{X_0} =0$.} We abuse terminology and refer to
    \begin{equation}
        X^{(n)} \to X_0^{(n)}
    \end{equation}
as a ``resolution'' of $X_0^{(n)}$, keeping in mind that our results can still be used to compute intersection numbers in models for which some singular fibers may remain over higher-codimension loci in $B^{(n - 1)}$, depending on the topology of $B^{(n - 1)}$. We furthermore assume that the blowups of $X_0^{(n)}$ can be realized explicitly as blowups of complete intersections of hyperplanes in the ambient space $Y_0^{(n + 1)}$ that intersect $X_0^{(n)}$ in singular subloci.

We associate to these blowups the blowdown maps
    \begin{equation}
         f_{i + 1}\colon Y^{(n + 1)}_{i + 1} \to Y^{(n + 1)}_i\,,
    \end{equation}
each of which contracts the exceptional divisor $ e_{i + 1} = 0$ in $Y_{i + 1}$. Given a sequence of $r$ blowups, the resolution $X^{(n)}$ is then a hypersurface (or more generally, a complete intersection) of an ambient projective bundle $Y^{(n + 1)}$, where $Y^{(n + 1)}$ is related to the ambient space $Y_0^{(n + 1)}$ of the original, singular elliptic fibration $X_0^{(n)}$ by a composition of blowdowns,
    \begin{equation}
        f_1 \circ \dotsb \circ f_{r - 1} \circ f_r\colon Y^{(n + 1)} \to Y_0^{(n + 1)}\,,
    \end{equation}
where, for simplicity, we write $Y^{(n + 1)} \equiv Y_r^{(n + 1)}$. In such cases, the canonical projection $\pi$ given in \cref{eq:canonicalprojection} can be lifted to the following composition of maps acting on the ambient space $Y^{(n + 1)}$,
    \begin{equation}
        \pi = \varpi \circ f_1 \circ \dotsb \circ f_{r - 1} \circ f_r\colon Y^{(n + 1)} \to B^{(n - 1)}\,,
    \end{equation}
so for the purposes of computing $\pi_*$, it suffices to compute the pushforward $f_{i *}$ defined with respect to each blowdown $f_i$ along with $\varpi_*$. The ambient projective bundle $Y^{(n + 1)}$ is in these cases equipped with a basis of divisors
    \begin{equation}
        \label{eq:Ydiv}
        \bm{D}_\alpha, \bm{H}, \bm{E}_i\,,
    \end{equation}
where $\bm{E}_i$ are the divisor classes associated to the total transforms of exceptional loci $e_i = 0$ contracted by the blowup maps $f_i$. Again, in well-behaved constructions, it is possible to represent the divisors $\hat{D}_I$ as restrictions of divisor classes $\hat{\bm{D}}_I$ in the Chow ring of $Y^{(n + 1)}$ to the hypersurface class $\bm{X}^{(n)}$, as in \cref{eq:hyprestrict}.

When the resolution $X^{(n)} \to X_0^{(n)}$ is given by a sequence of blowups as described above, it is possible to explicitly compute the pushforward map defined with respect to each blowdown $f_i$, where each pushforward map only acts non-trivially on the divisor class $\bm{E}_i$.

We now describe this procedure in more detail, namely the blowup of $Y_{i - 1}^{(n + 1)}$ along a complete intersection of hyperplanes. Let the fibers of $Y_{i - 1}^{(n + 1)}$ have homogeneous coordinates $y_{i - 1, a}$. We denote the center of the $i$th blowup by
    \begin{equation}
        g_i = (g_{i, 1}, g_{i, 2}, \dotsc, g_{i, n_i})\,, \quad g_{i, m} = g_{i, m}^a y_{i - 1, a}\,.
    \end{equation}
The generators $g_{i, m}$ are linear functions of the homogeneous coordinates of the fibers of the ambient projective bundle $Y_{i - 1}^{(n + 1)}$, so that the divisor classes associated to the vanishing loci of the generators of each blowup center $g_i$ given by (see \cref{foot:ydiv})
    \begin{equation}
        \label{eq:genclass}
        \bm{g}_{i, m} \equiv g^a_{i, m} \bm{y}_{i - 1, a} = g_{i, m}^a (y_{i - 1, a}^\alpha \bm{D}_\alpha +  y_{i - 1, a}^0 \bm{H} + y_{i - 1, a}^j \bm{E}_j)\,.
    \end{equation}
The blowup is carried out by making the formal substitutions\footnote{When the generators for the $i$th blowup are equal to one of the homogeneous coordinates, i.e., $g_{i, m} = y_{i - 1, a}$ for some $y_{i - 1, a}$ appearing in $[y_{i - 1, 1} : y_{i - 1, 2} : \dotsb]$, we abuse notation and write $g_{i, m} = y_{i - 1, a} \to e_i y_{i - 1, a}$.}
    \begin{equation}
       \label{eq:blowupsubs}
        g_{i, m} \to e_i y_{i, m}
    \end{equation}
and introducing as the exceptional locus $e_i = 0$ of the ambient space $Y_i^{(n + 1)}$ a new projective space with homogeneous coordinates
    \begin{equation}
        [y_{i, 1} : y_{i, 2}: \dotsb : y_{i, n_i}]\,.
    \end{equation}
To perform subsequent blowups, one simply iterates this procedure by collectively denoting \emph{all} of the homogeneous coordinates of the fibers of $Y_{i}^{(n + 1)}$ by $y_{i, a}$, and then repeating the formal substitutions \cref{eq:blowupsubs} for some collection of hyperplanes $g_{i + 1}$.

We now explain how to compute the pushforward defined with respect to an arbitrary blowdown map $f_i$. Given an arbitrary analytic function $F(\hat{D}_{\hat{I}})$ of the class $\hat{E}_i \equiv \bm{E}_i \cap \bm{X}^{(n)}$, we write
    \begin{equation}
        F(\hat{D}_{\hat{I}}) = F(\hat{\bm{D}}_{\hat{I}}) \cap \bm{X}^{(n)} \equiv \tilde{F}(\bm{E}_i).
    \end{equation}
The pushforward of $F(\hat{D}_{\hat{I}})$ to the Chow ring of $X_{i - 1}^{(n)}$ can be expressed in terms of a pushforward to the Chow ring of $Y_{i - 1}^{(n + 1)}$, given by\footnote{As before, we assume generically $\bm{g}_{i, m} \ne \bm{g}_{i, n}$ for $m \ne n$.}
    \begin{equation}
        \label{eq:singpush}
        f_{i *}(\tilde{F}(\bm{E}_i)) = \sum_{k = 1}^{n_i} \tilde{F}(\bm{g}_{i, k})  \prod_{\substack{m = 1 \\ m \ne k}}^{n_i} \frac{\bm{g}_{i, m}}{\bm{g}_{i, m} - \bm{g}_{i, k}}\,.
    \end{equation}
This implies that if we again write an arbitrary analytic function $F(\hat{D}_{\hat{I}})$ of the divisor classes $\hat{D}_{\hat{I}}$ as $ \tilde{F}(\bm{E}_i)$, as in \cref{eq:analyticlift} (where now the basis of divisors of $Y^{(n + 1)}$ also includes the classes $\bm{E}_i$ of the total transforms of the exceptional loci in the ambient projective bundle that are contracted by each blowdown), the pushforward defined with respect to the map $\pi\colon X^{(n)} \to X_0$ is given by the composition
    \begin{equation}
        \label{eq:pushcomp}
        \pi_*(F(\hat{D}_{\hat{I}})) = \varpi_* \circ f_{1 *} \circ \dotsb \circ f_{r *}(\tilde{F}(\bm{E}_r))\,.
    \end{equation}

\subsubsection{Example: blowups of the projective plane}

We illustrate the formula \labelcref{eq:singpush} with a simple example, namely a nodal curve in $Y_0^{(2)} = \bP^2$ with homogeneous coordinates $[x : y : z]$:
    \begin{equation}
        X_0^{(1)}\colon -y^2 z + x^3 + x^2 z = 0\,.
    \end{equation}
(Note that the nodal curve is not a fibration. We do not require this example to be a fibration, as our goal is simply to illustrate the action of a pushforward map defined with respect to a blowdown.) The curve $X_0^{(1)}$, which is a degree-three hypersurface of $Y_0^{(2)}$, has a singularity along $x = y = 0$. We resolve this singularity by introducing a blowup along the locus $g_{1, 1} = g_{1, 2} = 0$ with
    \begin{equation}
        g_{1,1} = x\,, \quad g_{1, 2} = y\,.
    \end{equation}
That is, we make the replacements $x \to e_1 x, y \to e_1 y$ so that the (smooth) proper transform is given by
    \begin{equation}
        X^{(1)}\colon -y^2 z + e_1 x^3 + x^2 z = 0
    \end{equation}
in the ambient space $Y^{(2)} \cong \operatorname{Bl}_1 \bP^2$ with homogeneous coordinates $[e_1 x : e_1 y : z] [x : y]$. The ambient space has a basis of divisors
    \begin{equation}
        \bm{H}, \bm{E}_1\,,
    \end{equation}
where $\bm{H}$ is the hyperplane class of $Y_0^{(2)} = \bP^2$ and $\bm{E}_1$ is the exceptional divisor $e_1 = 0$ of the blowup contracted by the map
    \begin{equation}
        f_1\colon Y^{(2)} \to Y_0^{(2)}\,.
    \end{equation}
Likewise, we can write down a basis of divisors of $X^{(1)}$. For example, the class of the intersection of the hyperplane $x = 0$ with $X^{(1)}$ is
    \begin{equation}
    (\bm{H} - \bm{E}_1) \cap \bm{X}^{(1)} = (\bm{H} - \bm{E}_1) (3 \bm{H} - 2 \bm{E}_1)\,.
    \end{equation}
In order to evaluate the right-hand side of this expression, we need to compute the intersection numbers of the ambient space. This computation can be done using \cref{eq:singpush}. We find:
    \begin{equation}
    \begin{aligned}
        f_{1 *}(\bm{E}_1) &= \bm{x} \frac{\bm{y}}{\bm{y} - \bm{x}} + \bm{y} \frac{\bm{x}}{\bm{x} - \bm{y}} = 0\,, \\
        f_{1 *}(\bm{E}_1^2) &= \bm{x}^2 \frac{\bm{y}}{\bm{y} - \bm{x}} + \bm{y}^2 \frac{\bm{x}}{\bm{x} - \bm{y}} = \bm{x} \bm{y} \left(\frac{\bm{x}}{\bm{y} - \bm{x}} + \frac{\bm{y}}{\bm{x} - \bm{y}}\right) = -\bm{x} \bm{y} = -\bm{H}^2 = -1\,.
    \end{aligned}
    \end{equation}
The above computation simply recovers the fact that $\bm{E}_1$ is an exceptional curve, with self-intersection $-1$. Thus, we learn that
    \begin{equation}
        (\bm{H} - \bm{E}_1) (3 \bm{H} - 2 \bm{E}_1) = 3 \bm{H}^2 + 2 \bm{E}_1^2 = 1
    \end{equation}
where above we have used the fact that $\bm{H} \bm{E}_1 = 0$ since the pushforward acts trivially on $\bm{H}$ and the pushforward of $\bm{E}_1$ is 0. Thus, we find that the locus $x = -y^2 z + e_1 x^3 + x^2 z = 0$ in $X^{(1)}$ consists of a single point.

\section{Computational algorithm}
\label{sec:alg}

We now present the central result of this paper, which is a simple computational algorithm designed to significantly speed up the evaluation of the pushforwards of topological intersection numbers of elliptic $n$-folds $X^{(n)}$ defined over an arbitrary smooth base $B^{(n - 1)}$ of complex dimension $n - 1$. This result applies the pushforward formulas \cref{eq:smoothpush,eq:singpush} to the special case where $X^{(n)}$ is a resolution of a singular elliptic $n$-fold $X_0^{(n)}$. Our algorithm can be summarized by the following steps:
    \begin{itemize}
        \item \textbf{Define a generating function} The first step is trivial. We simply define the generating function
            \begin{equation}
                e^{\hat{J}}\,, \quad \hat{J} = \phi^I \hat{D}_I\,,
            \end{equation}
       where $\phi^I$ may be regarded as real-valued K\"ahler parameters. The purpose of this step is to identify a convenient analytic function that encodes the intersection numbers, so that we can eliminate the computational expense involved in computing the pushforwards of individual intersection numbers, and simply compute the pushforward of $e^{\hat{J}}$. The reason that this strategy works is that the pushforward map acts linearly on products of divisor classes and can hence be applied to any analytic function $F(\hat{D}_I)$ that admits a power series expansion. In particular, the power series expansion of the generating function $e^{\hat{J}}$ encodes all intersection products as its summands.

        This strategy was used in~\cite{Esole:2017kyr} to succinctly compute the generating function of Euler characteristics for $\G$-models defined over an arbitrary smooth base $B^{(n - 1)}$, where the Euler characteristic was recovered as the top Chern class, i.e., the degree-$n$ element of the pushforward of the total Chern class. One important computational advantage in this case is that $e^{\hat{J}}$ is a simpler symbolic expression than the total Chern class, and leads to fewer spurious poles that require careful cancellation.

        \item \textbf{Pre-define an array of dummy functions} The second step is to actually evaluate the pushforward map explicitly. As we have described, we focus on elliptic $n$-folds $X^{(n)}$ that resolve singular elliptic $n$-folds via a sequence of blowups of the ambient space along complete intersections of hyperplanes. In such cases, the (lift of the) canonical projection $\pi\colon X^{(n)} \to B^{(n - 1)}$ can be expressed as the composition $\pi = \varpi \circ f_1 \circ \cdots \circ f_r$ and hence the pushforward $\pi_*$ is given by \cref{eq:pushcomp}. Evaluating the action of the pushforward map on the non-trivial part of the generating function, $e^{\phi^{\hat{I}} \hat{D}_{\hat{I}}}$, allows us to compute an analytic function
            \begin{equation}
                \cZ_\phi(L_a, \Sigma_s) = \pi_*(e^{\phi^{\hat{I}} \hat{D}_{\hat{I}}})
            \end{equation}
        whose derivatives with respect to $\phi^{\hat{I}}$ capture the intersection numbers of $X^{(n)}$ in terms of intersection products involving the characteristic divisor classes $L_a$ and relevant codimension-one components $\Sigma_s$ of the discriminant locus of $X_0^{(n)}$ in $B^{(n - 1)}$. The full generating function of intersection numbers is $Z_\phi = \pi_*(e^{\hat{J}}) = e^{\phi^\alpha D_\alpha} \cZ_\phi$.

        The novel part of this algorithm is the actual method we use to evaluate $\pi_*$. The idea is to pre-define a collection of substitution functions in a symbolic computing program such as \emph{Mathematica}, whose composition can then be applied to $e^{\hat{J}}$. More concretely, suppose we start with a resolution $X^{(n)} \to X_0^{(n)}$ consisting of $r$ blowups whose generators $\bm{g}_i$ are all known; see \cref{eq:genclass}. This means that the pushforward map $\pi_*$ will be a composition of $r + 1$ maps,
            \begin{equation}
                \label{eq:algcomp}
                \pi_* =  \varpi_* \circ f_{1 } \circ \dotsb \circ f_{r *}\,.
            \end{equation}
        We use the basis $\bm{D}_\alpha, \bm{H}, \bm{E}_i$ for the ambient space. Assume that a basis of divisors $\hat{D}_I$ is known and can be expressed as the restrictions of linear combinations of the ambient space divisors restricted to the class of the hypersurface. Suppose we want to compute the pushforward of an arbitrary function
            \begin{equation}
                F(\bm{D}_\alpha, \bm{H}, \bm{E}_i) \cap \bm{X}^{(n)} \equiv \tilde{F}(\bm{E}_i)\,,
            \end{equation}
        where on the right-hand side of the above equation, to start, we simply treat $\tilde{F}$ as a function of the class $\bm{E}_i$. We can create a template of dummy functions to represent the pushforward under each element of the composition in \cref{eq:algcomp}. For example, the pushforward
            \begin{equation}
                f_{r *}(\tilde{F}(\bm{E}_r)) = \sum_{k_r = 1}^{n_r} \tilde{F}(\bm{g}_{r, k_r}) \prod_{\substack{m = 1 \\ m \ne k_r}}^{n_r} \frac{\bm{g}_{r, m}}{\bm{g}_{r, m} - \bm{g}_{r, k_r}} \equiv \sum_{k_r =1}^{n_r} \tilde{F}_{k_r}(\bm{E}_{r - 1})
            \end{equation}
        can be pre-defined, where we note that the classes $\bm{g}_{r, k}$ are treated as input since we are assuming that the resolution is known ahead of time. Note that we have introduced a new definition on the right-hand side of the above equation. By linearity, this implies that the second pushforward, computed with respect to $f_{r - 1}$, can be pre-defined:
            \begin{equation}
            \begin{aligned}
                f_{r - 1 *}(\sum_{k_r = 1}^{n_r} \tilde{F}_{k_r}(\bm{E}_{r - 1})) &= \sum_{k_r, k_{r - 1} = 1}^{n_r, n_{r - 1}} \tilde{F}_k(\bm{g}_{r - 1, k_{r - 1}}) \prod_{\substack{m = 1 \\ m \ne k_{r - 1}}}^{n_{r - 1}} \frac{\bm{g}_{r - 1, m}}{\bm{g}_{r - 1, m}- \bm{g}_{r - 1, k_{r - 1}}} \\
                &\equiv \sum_{k_r, k_{r - 1} = 1}^{n_r, n_{r - 1}} \tilde{F}_{k_r, k_{r - 1}}(\bm{E}_{r - 2})\,,
             \end{aligned}
            \end{equation}
        where we have again introduced a new definition on the right-hand side of the above equation. Proceeding in this manner, we can continue pre-defining functions for all partial compositions of the pushforwards, until we arrive at an expression
            \begin{equation}
                f_{1 *} \circ f_{2 *} \circ \dotsb \circ f_{r - 1 *} \circ f_{r *}(\tilde{F}(\bm{E}_r)) \equiv \sum_{k_r, k_{r - 1}, \dotsc, k_1 = 1}^{n_r,  n_{r - 1}, \dotsc, n_1} \tilde{F}_{k_r, k_{r - 1}, \dotsc, k_1}(\bm{H})\,.
            \end{equation}
        Computing the pushforward of the above function with respect to $\varpi$, we obtain
            \begin{equation}
                 \varpi_* \circ f_{1 *} \circ f_{2 *} \circ \dotsb \circ f_{r - 1 *} \circ f_{r *}(\tilde{F}(\bm{E}_r)) = \sum_{a = 1}^3 \frac{\sum_{k_r, k_{r - 1}, \dotsc, k_1 = 1}^{n_r, n_{r - 1}, \dotsc, n_1} \tilde{F}_{k_r, k_{r - 1}, \dotsc, k_1}(-L_a)}{\prod_{b\ne a} (L_a - L_b)}\,.
            \end{equation}
        To summarize, this procedure has enabled us to symbolically pre-define the following set of dummy functions:
            \begin{equation}
            \begin{aligned}
                \label{eq:dummy}
                \{f_{r *}(\tilde{F}(\bm{E}_r)), \dotsc, &\varpi_* \circ f_{1 *} \circ \dotsb \circ f_{r *}(\bm{E}_r)\} \\
                &= \left\{\sum_{k_r = 1}^{n_r} \tilde{F}_{k_r}(\bm{E}_{r-1}), \dotsc, \sum_{a = 1}^3 \frac{\sum_{k_r, k_{r - 1}, \dotsc, k_1}^{n_r, n_{r - 1}, \dotsc, n_1} \tilde{F}_{k_r, k_{r - 1}, \dotsc, k_1}(-L_a)}{\prod_{b \ne a} (L_a - L_b)}\right\}\,.
            \end{aligned}
            \end{equation}
        \item \textbf{Substitute the generating function for the first dummy function} The power of this algorithm is that we can define the full set of dummy functions \labelcref{eq:dummy} at the beginning of the computation, thereby reducing the remainder of the computation to a chain of symbolic substitutions. To complete the computation, we must supply a specific function $\tilde{F}$ (such as the generating function $e^{\hat{J}}$) as input, and then recursively convert the remaining dummy functions into concrete pushforwards by way of direct substitution. The symbolic substitutions required to compute these pushforwards are not necessarily simple because complicated spurious poles can emerge, and need to be carefully cancelled. However, provided one can identify a suitable means to cancel these spurious poles, the substitutions can be carried out straightforwardly and an explicit expression for the pushforward
            \begin{equation}
                \cF(L_a, \Sigma_s) = \varpi_* \circ f_{1 *} \circ \dotsb \circ f_{r *}(F(\bm{D}_\alpha, \bm{H}, \bm{E}_i) \cap \bm{X}^{(n)})
            \end{equation}
        can be obtained. Note that as of the time of writing, the current version of \emph{Mathematica} has two functions \texttt{Apart[]} and \texttt{Cancel[]} that can be used to eliminate these spurious poles at each step of the substitution.
    \end{itemize}
We have implemented the above algorithm in a \emph{Mathematica} package \texttt{IntersectionNumbers.m}. The reader interested in immediately using the function \texttt{push[]} to compute pushforwards should see \cref{sec:main} for a quick start guide and refer to the documentation included in \texttt{IntersectionNumbers.m} for additional information.

\section{Generating functions of intersection numbers for F-theory Tate models}
\label{sec:generating}

\subsection{Preliminaries}
In this section, we describe how to compute the pushforward of a particular analytic function of the divisors $\hat{D}_I$ of a smooth elliptically fibered $n$-fold
    \begin{equation}
        \pi\colon X^{(n)} \to B^{(n - 1)}\,,
    \end{equation}
namely the generating function
\begin{equation}
        e^{\hat{J}}\,, \quad \hat{J} = \phi^I \hat{D}_I
    \end{equation}
of $k$-fold intersection products $\hat{D}_{I_1} \hat{D}_{I_2} \dotsm \hat{D}_{I_{k - 1}} \hat{D}_{I_k}$, see \cref{eq:derivative}. When $k = n$, the above intersection product (computed in the Chow ring of $X^{(n)}$) is a number; when $k < n$, the above intersection product corresponds to a degree-$k$ element of the Chow ring whose intersection product with a degree $n - k$ element produces a number; finally, when $k > n$, the above intersection product vanishes identically, although such an element can be interpreted as a non-trivial intersection product in the Chow ring of a higher-dimensional elliptic fibration by formally extending the complex dimension of the base to be $n' - 1 \ge k - 1 > n - 1$. In the spirit of this formal extension, if we take $\hat{J} = \phi^I \hat{D}_I$ to be the K\"ahler class of an elliptically fibered $n$-fold, the generating function can be interpreted as a formal infinite sum of the volumes
    \begin{equation}
        e^{\hat{J}} = \sum_{n = 0}^\infty \vol(X^{(n)})\,, \quad \vol(X^{(n)}) = \frac{1}{n!} \hat{J}^n
    \end{equation}
corresponding to bases of incrementally increasing complex dimension $n - 1$.

When $X^{(n)}$ is in particular a resolution of a singular Weierstrass model $X_0^{(n)}$ defining a F-theory model with gauge group $\G$, the Shioda--Tate--Wazir formula implies
    \begin{equation}
        h^{1, 1}(X^{(n)}) = 1 + h^{1, 1}(B^{(n - 1)}) + \rank(\G)\,,
    \end{equation}
and hence there exists a canonical basis of (not necessarily primitive\footnote{We use the non-standard terminology ``primitive divisor'' to mean a divisor whose Chow ring class cannot be expressed as a non-trivial positive linear combination of the classes of other effective divisors.}) divisors
    \begin{equation}
        \label{eq:ellipticbasis}
        \hat{D}_I =  \hat{D}_\alpha, \hat{D}_0, \hat{D}_i\,.
    \end{equation}
Here, $\hat{D}_0$ is the zero section, $\hat{D}_\alpha$ are pullbacks of divisors on $B^{(n - 1)}$, and $\hat{D}_i$ are divisors associated to Cartan elements of the simple non-abelian gauge factors $\G_s \subset \G$.\footnote{In this paper, we assume the free abelian part of $\G$ is empty and hence that $\G$ is strictly non-abelian.}

Since the pushforward map $\pi_*$ only acts non-trivially on the divisors $\hat{D}_0, \hat{D}_i$ in these models, the final expression for the pushforward of the generating function factorizes as follows:
    \begin{equation}
        \label{eq:genZnoa}
        Z_\phi = \pi_*(e^{\phi^\alpha \hat{D}_\alpha + \phi^0 \hat{D}_0 + \phi^i \hat{D}_i}) = e^{\phi^\alpha D_\alpha} \cZ_\phi
    \end{equation}
where
    \begin{equation}
        \cZ_\phi = \varpi_* \circ f_{1 *} \circ \dotsb \circ f_{k *}(e^{\phi^0 \hat{\bm{D}}_0 + \phi^i \hat{\bm{D}}_i} \cap \bm{X}^{(n)})\,.
    \end{equation}
In the remainder of this section, we explain in detail how to compute $Z_\phi$ explicitly for a standard set of resolutions of various F-theory Tate models.

\subsection{Examples of F-theory Tate models}
\label{eq:Tate}

In the following subsections, we describe standard resolutions for F-theory Tate models with gauge groups $\G = \SU(r + 1), \SO(2 r + 1), \Sp(r), \SO(2 r)$ of arbitrary rank $r$, along with explicit computations of the pushforward of the generating function $e^{\hat{J}}$ for some examples of models with small $r$. Examples of pushforwards of generating functions of intersection numbers for groups with larger $r$ are collected in an ancillary \emph{Mathematica} notebook accompanying the upload of this preprint to the arXiv.

The F-theory Tate models we describe below are each defined by a singular elliptic fibration $X_0^{(n)}$ constructed explicitly as the zero locus of a homogeneous polynomial:
    \begin{equation}
        \label{eq:Tateform}
        y^2 z + a_1 x y z + a_3 y z^2 - (x^3 + a_2 x^2 z + a_4 x z^2 + a_6 z^3) = 0\,.
    \end{equation}
The above equation is the most general version of the Weierstrass equation \cref{eq:Weierstrass} as written in the notation of Tate~\cite{10.1007/BFb0097582}. A singular elliptic fibration defined by a hypersurface equation of the above form is sometimes referred to as a ``Tate model'' in the F-theory literature. One particularly useful property of Tate models is that the specification of the singularity types of the elliptic fiber over codimension-one components of the discriminant locus $\Delta = 0$ in terms of orders of vanishing of $f, g, \Delta$ can be further refined into a specification of the orders of vanishing of the sections $a_m$. The orders of vanishing of $a_m$ can be used to control the monodromy of $f, g, \Delta$, and hence can distinguish models with simply laced gauge algebra from models with non-simply laced gauge algebra $\Lie(\G)$; see \cref{tab:Tatetable}. The full specification of the F-theory gauge algebra $\Lie(\G)$ in terms of the orders of vanishing of $a_m$ in the F-theory literature is viewed as an application of Tate's algorithm~\cite{10.1007/BFb0097582} (we sometimes refer to this as ``Tate tuning'') to the geometry of singular elliptic fibrations. In particular, the F-theory Tate models we study in this section are each defined by a specific choice of Tate tuning, as we discuss in more detail below.
\begin{table}
    \centering

        $\begin{array}{|c|c|c|c|c|c|c|c|}\hline
            \text{fiber type} & \ord(a_1) & \ord(a_2) & \ord(a_3) & \ord(a_4) & \ord(a_6) & \ord(\Delta) & \Lie(\G) \\\hline \hline
            \singtype{I}_0 & 0 & 0 & 0 & 0 & 0 & 0 & \text{---} \\\hline
            \singtype{I}_1 & 0 & 0 & 1 & 1 & 1 & 1 & \text{---} \\\hline
            \singtype{I}_2 & 0 & 0 & 1 & 1 & 2 & 2 &\gA_1 \\\hline
            \singtype{I}_3^{\text{ns}} & 0 & 0 & 2 & 2 & 3 & 3 & \gC_1 \\\hline
            \singtype{I}_3^{\text{s}}& 0 & 1 & 1 & 2 & 3 & 3 & \gA_2 \\\hline
            \singtype{I}_{2 n}^{\text{ns}} & 0 & 0 & n & n & 2 n & 2 n & \gC_n \\\hline
            \singtype{I}_{2 n}^{\text{s}} & 0 & 1 & n & n & 2 n & 2 n & \gA_{2 n - 1} \\\hline
            \singtype{I}_{2 n}^{\text{s}} \text{ (2nd ver.)} & 0 & 2 & n-1 & n + 1 & 2 n & 2 n & \gA_{2 n - 1}^\circ \\\hline
            \singtype{I}_{2 n + 1}^{\text{ns}} & 0 & 0 & n + 1 & n + 1 & 2 n + 1 & 2 n + 1 & \gC_n \\\hline
            \singtype{I}_{2 n + 1}^{\text{s}} & 0 & 1 & n & n + 1 & 2 n + 1 & 2 n + 1 & \gA_{2 n} \\\hline
            \singtype{II} & 1 & 1 & 1 & 1 & 1 & 2 & \text{---} \\\hline
            \singtype{III} & 1 & 1 & 1 & 1 & 2 & 3 & \gA_2 \\\hline
            \singtype{IV}^{\text{ns}} & 1 & 1 & 1 & 2 & 2 & 4 & \gC_1 \\\hline
            \singtype{IV}^{\text{s}} & 1 & 1 & 1& 2 & 3 \, (2)^\dagger &4 & \gA_2 \\\hline
            \singtype{I}_0^{* \text{ns}} & 1 & 1 & 2 & 2 & 3 & 6 & \gG_2 \\\hline
            \singtype{I}_0^{*\text{ss}} & 1 & 1 & 2 & 2 & 4 & 6 & \gB_3 \\\hline
            \singtype{I}_0^{* \text{s}} & 1 & 1 & 2 & 2 & (4,3)^\dagger & 6 & \gD_4 \\\hline
            \singtype{I}_{2 n - 3}^{* \text{ns}} & 1 & 1 & n & n + 1 & 2 n & 2 n + 3 & \gB_{2 n} \\\hline
            \singtype{I}_{2 n - 3}^{* \text{s}} & 1& 1 & n & n + 1 & 2 n + 1 \, (2 n)^\dagger & 2 n + 3 & \gD_{2 n + 1} \\\hline
            \singtype{I}_{2 n - 2}^{* \text{ns}} & 1 & 1 & n + 1 & n + 1 & 2 n + 1 & 2 n + 4 & \gB_{2 n + 1} \\\hline
            \singtype{I}_{2 n - 2}^{* \text{s}} & 1 & 1 & n + 1 & n + 1 & 2 n + 2 \, (2 n + 1)^\dagger & 2 n + 4 & \gD_{2 n + 2} \\\hline
            \singtype{IV}^{* \text{ns}} & 1 & 2 & 2 & 3 & 4 & 8 & \gF_4 \\\hline
            \singtype{IV}^{* \text{s}} & 1 & 2 & 2 & 3 & 5 \, (4)^\dagger & 8 & \gE_6 \\\hline
            \singtype{III}^* & 1 & 2 & 3 & 3 & 5 & 9 & \gE_7 \\\hline
            \singtype{II}^* & 1 & 2 & 3 & 4 & 5 & 10 & \gE_8 \\\hline
            \text{non-min.} & 1 & 2 & 3 & 4 & 6 & 12 & \text{?} \\\hline
        \end{array}$

    \caption{This table, which is a reproduction of Table~4 in~\cite{Huang:2018gpl}, lists the orders of vanishing of the sections $a_m$ multiplying the monomials appearing in the Tate form of the Weierstrass equation appearing in \cref{eq:Tateform}. The superscripts $\text{`ns'}, \text{`ss'}, \text{`s'}$ in the first column are short for (resp.) `non-split', `semi-split', and `split'; these superscripts specify the arithmetic geometric properties of the Kodaira fibers, and in particular indicate whether or not the monodromy of the sections $f, g$ around the corresponding codimension-one components of the discriminant locus lead to a non-simply laced version of the gauge algebra in the low-energy effective description of the F-theory compactification. Orders of vanishing marked with a $\dagger$ superscript require additional monodromy conditions in order to realize the gauge algebra listed in the same row. Note that the $\gA_{2 n - 1}^\circ$ F-theory model includes exotic matter content in the case $n = 3$.}
    \label{tab:Tatetable}
\end{table}

\Cref{eq:Tateform} defines a hypersurface of the ambient projective bundle $Y^{(n + 1)} = \bP(\sV) \to B^{(n - 1)}$ with $\sV = \sL^{\otimes 2} \oplus \sL^{\otimes 3} \oplus \sO$, where $\sL \to B$ is a line bundle characterizing the elliptic fibration (recall that the CY condition is imposed by setting $\sL$ equal to the anticanonical bundle of $B^{(n - 1)}$; see the discussion around \cref{eq:CYcondition}). The fibers of $Y^{(n + 1)}$ are isomorphic to $\bP^2$ with homogeneous coordinates $[x : y : z]$, while the parameters $a_m$ are sections of various tensor powers of the line bundle $\sL = \sK^{-1}$. The classes of the divisors corresponding to the zero loci of the variables appearing in \cref{eq:Tateform} are
\begin{equation}
        \bm{x} = \bm{H} - 2 \bm{K}\,, \quad \bm{y} = \bm{H} - 3 \bm{K}\,, \quad \bm{z} = \bm{H}\,, \quad \bm{a}_m = -m \bm{K}\,,
    \end{equation}
which implies that the divisor class of the hypersurface $X_0^{(n)}$ is
    \begin{equation}
        \bm{X}_0^{(n)} = 3 \bm{H} - 6 \bm{K}\,.
    \end{equation}
The elliptic fibration $X_0^{(n)}$ is not in general smooth, and the singular fibers of $X_0^{(n)}$ can be identified as the fibers over the discriminant locus $\Delta = 0$ (in $B^{(n - 1)}$) of the polynomial in \cref{eq:Tateform}\footnote{Note that the discriminant $\Delta$ of the polynomial in \cref{eq:Tateform} matches the discriminant of the Weierstrass polynomial $y^2 - (x^3 + f x + g)$, namely $\Delta = 4 f^3 + 27 g^2$.}. Following Tate's algorithm, it is possible to prescribe the singularity type of elliptic fibers over an irreducible codimension-one component $\Sigma$ of the discriminant locus by requiring certain orders of vanishing for the sections $a_m$ in a neighborhood of $\Sigma$. More specifically, if we describe $\Sigma$ by the local equation $\sigma = 0$ and expand the sections $a_m$ in a local power series in the vicinity of $\sigma$ as
    \begin{equation}
        \label{eq:Tatetuning}
        a_m = a_{m, n_m} \sigma^{n_m}\,,
    \end{equation}
specifying the orders of vanishing of $a_m$ corresponds to picking the smallest values of $n_m$ for which $a_{m, n_m}$ are non-vanishing. Note that in the Chow ring of $B$, the divisor classes of the sections $a_{m, n_m} = 0$ are given by
    \begin{equation}
        [a_{m, n_m} = 0] = -m K - n_m \Sigma\,, \quad \Sigma = [\sigma = 0]\,, \quad -K = c_1(\sK^{-1})\,,
    \end{equation}
and as usual, we denote the pullbacks of the above classes to the Chow ring of $Y^{(n + 1)}$ by (resp.) $\bm{a}_{m, n_m} = -m \bm{K} - n_m \bm{\Sigma}$ and $\bm{\sigma} = \bm{\Sigma}$.

In order to resolve the singularities of $X_0^{(n)}$, we follow the procedure described in \cref{sec:pushsing} and introduce a sequence of blowups of various loci in the ambient space that restrict to components of the singular locus of the hypersurface. Given a Tate tuning corresponding to an F-theory model with gauge symmetry group $\G$, we introduce a sequence of $r = \rank(\G)$ blowups, where we denote the class of the exceptional divisor of the $i$th blowup by $\bm{E}_i$, leading to a basis of divisors of the ambient projective bundle $Y_r^{(n + 1)}$ of the form given in \cref{eq:Ydiv}.

For convenience, we parametrize the generating function in \cref{eq:genZnoa} in the $\bm{E}_i$ basis, using the fact that the Cartan divisors can be written as linear functions
    \begin{equation}
        \hat{\bm{D}}_i  = \ell_i^j \bm{E}_j\,,
    \end{equation}
so that
    \begin{equation}
    \begin{aligned}
        e^{\hat{J}} &= e^{\phi^\alpha \hat{\bm{D}}_\alpha} e^{\phi^0 \hat{\bm{D}}_0 + \phi^i \hat{\bm{D}}_i} \cap \bm{X}^{(n)} \\
        &= e^{\phi^\alpha \hat{\bm{D}}_\alpha} e^{\frac{1}{3} \alpha^0 \bm{H} + \alpha^i \bm{E}_i} (3 \bm{H} - 6 \bm{K} - \sum_i (n_i - 1) \bm{E}_i)\,,
    \end{aligned}
    \end{equation}
where $n_i$ is the number of generators determining the center of the $i$th blowup. We compute the pushforward of the above expression to the Chow ring of $B^{(n - 1)}$, which can be expressed as an analytic function of the real parameters $\alpha^0, \alpha^i$, with coefficients given by intersection products of the divisors $K, \Sigma$. The change of basis from the $\bm{E}_i$ basis (parametrized by $\alpha^i$) to the basis of Cartan divisors $\hat{\bm{D}}_i$ (parametrized by the K\"ahler parameters $\phi^i$) can be implemented by the change of coordinates
    \begin{equation}
        \alpha^i = \phi^j \ell_j^i\,.
    \end{equation}
One simple check that the change of basis is implemented correctly is that the matrix of double intersection numbers of Cartan divisors reproduces the inverse metric tensor $\kappa_{i j}$ of the Lie algebra $\Lie(\G)$:
    \begin{equation}
        \label{eq:Cartan}
        \kappa_{i j} = \left.\frac{\partial}{\partial\phi^i} \frac{\partial}{\partial\phi^j} Z_\phi\right|_{\phi^I = 0}\,.
    \end{equation}

\subsubsection{$\SU(r + 1)$ model}
\label{sec:SUr+1model}

Assume $r > 1$. The $\SU(r + 1)$ Tate model is characterized by a $\singtype{I}_{r + 1}^\text{split}$ singularity over a divisor $\Sigma \subset B^{(n - 1)}$, and can be constructed via the Tate tuning
    \begin{equation}
        a_1 = a_1\,, \quad a_2 = a_{2, 1} \sigma\,, \quad a_3 = a_{3, \floor*{\frac{r + 1}{2}}} \sigma^{\floor*{\frac{r + 1}{2}}}\,, \quad a_4 = a_{4, \ceil*{\frac{r + 1}{2}}} \sigma^{\ceil*{\frac{r + 1}{2}}}\,, \quad a_6 = a_{6, r + 1} \sigma^{r + 1}\,.
    \end{equation}
We consider the family of resolutions $X_r^{(n)} \to X_0^{(n)}$ given by the following sequence of blowups~\cite{Esole:2015xfa}:
    \begin{equation}
        \label{eq:Ares}
        X_r^{(n)} \xrightarrow{(*, e_{r - 1} | e_r)} \dotsb \xrightarrow{(x, e_2 | e_3)} X_2^{(n)} \xrightarrow{(y, e_1 | e_2)} X_1^{(n)} \xrightarrow{(x, y, \sigma | e_1)} X_0^{(n)}\,, \quad * = \begin{cases} x, & r \text{ odd} \\ y, & r \text{ even}
        \end{cases}\,.
    \end{equation}
The change of basis from the $\bm{E}_i$ basis to the basis of Cartan divisors is given by
    \begin{equation}
        \alpha^i = (-1)^{i + 1} (\phi^{\ceil*{\frac{i}{2}}} - \phi^{r + 1 - \floor*{\frac{i}{2}}})\,, \quad \phi^{r + 1} \equiv 0\,.
    \end{equation}

\subsubsection{A detailed example: $\SU(2)$}

As a detailed example of the previous section, consider the case $r = 2$, i.e., the $\SU(2)$ model over an arbitrary smooth base $B^{(n - 1)}$. Since the rank of the gauge group is minimal in this case, the algebraic manipulations involved in evaluating the pushforward are not prohibitively complicated and hence we use this example as an opportunity to spell out the details.

We consider the resolution
    \begin{equation}
        \label{eq:SU2blowup}
        X_1^{(n)} \xrightarrow{(x, y, \sigma | e_1)} X_0^{(n)}\,,
    \end{equation}
where the proper transform $X_1^{(n)} \subset Y_1^{(n + 1)}$ is given by the hypersurface equation
    \begin{equation}
        y^2 z + a_1 x y z +  a_{3, 1} \sigma y z^2 - (e_1 x^3 + a_2 x^2 z + a_{4, 1} \sigma x z^2 + a_{6, 2} \sigma^2 z^3) = 0
    \end{equation}
in the ambient space $Y_1^{(n + 1)}$ whose fibers are equipped with homogeneous coordinates $[x e_1 : y e_1 : z] [x : y : \sigma]$. The equation $e_1 = 0$ is a local equation for the exceptional divisor in $Y_1^{(n + 1)}$. The divisor classes of the generators $x, y, \sigma$ of the blowup in \cref{eq:SU2blowup} are
    \begin{equation}
        \label{eq:SU2blowupgen}
        \bm{x} = \bm{H} - 2 \bm{K}\,, \quad \bm{y} = \bm{H} - 3 \bm{K}\,, \quad \bm{\sigma} = \bm{\Sigma}\,,
    \end{equation}
and the class of the exceptional divisor is given by
    \begin{equation}
        [e_1 = 0] = \bm{E}_1\,.
    \end{equation}
We choose a basis of divisors $\hat{D}_\alpha, \hat{D}_0, \hat{D}_1$ for $X_1^{(n)}$ and, similarly, a basis of divisors $\bm{H}, \bm{D}_\alpha, \bm{E}_1$ for $Y_1^{(n + 1)}$. The relationship between these two bases of divisors is given by
    \begin{equation}
        \hat{D}_0 = \frac{1}{3} \bm{H} \cap \bm{X}_1^{(n)}\,, \quad \hat{D}_\alpha = \bm{D}_\alpha \cap \bm{X}_1^{(n)}\,, \quad \hat{D}_1 = \bm{E} \cap \bm{X}_1^{(n)}\,,
    \end{equation}
where
    \begin{equation}
        \bm{X}_1^{(n)} = \bm{X}_0^{(n)} - 2 \bm{E}_1 = 3 \bm{H} - 6 \bm{K} - 2 \bm{E}_1
    \end{equation}
is the divisor class of the hypersurface $X_1^{(n)} \subset Y_1^{(n + 1)}$; this is simply a reflection of the fact that the divisors of the ambient space restrict to divisors of the hypersurface in this construction.

Our next step is to write the generating function for the intersection numbers of the above divisors:
    \begin{equation}
        \label{eq:SU2genfun}
        e^{\hat{J}} =  e^{\phi^\alpha \hat{\bm{D}}_\alpha} e^{\frac{\alpha^0}{3} \bm{H} + \alpha^1 \bm{E}_1} \cap \bm{X}_1^{(n)}\,.
    \end{equation}
We denote the blowdown map that contracts the exception divisor located at $e_1 = 0$ by $f_1$, and as usual we denote the canonical projection $Y_0^{(n + 1)} \to B^{(n - 1)}$ by $\varpi$. In order to evaluate the pushforward of the above expression with respect to the composition of maps $\varpi \circ f_1$, we recall (\cref{eq:singpush}) the action of the pushforward map $f_{1 *}$ on an arbitrary analytic function of the class $\bm{E}_1$:
    \begin{equation}
    \begin{aligned}
        f_{1 *}(\tilde{F}(\bm{E}_1)) &= \sum_{k = 1}^3 \tilde{F}(\bm{g}_{i, k}) \prod_{\substack{m = 1 \\ m \ne k}}^3 \frac{\bm{g}_{i, m}}{\bm{g}_{i, m} - \bm{g}_{i, k}} \\
        &= \tilde{F}(\bm{x}) \frac{\bm{y}}{\bm{y} - \bm{x}} \frac{\bm{\sigma}}{\bm{\sigma} - \bm{x}} + \tilde{F}(\bm{y}) \frac{\bm{\sigma}}{\bm{\sigma} - \bm{y}} \frac{\bm{x}}{\bm{x} - \bm{y}} + \tilde{F}(\bm{\sigma}) \frac{\bm{x}}{\bm{x} - \bm{\sigma}} \frac{\bm{y}}{\bm{y} - \bm{\sigma}} \\
        &\equiv \sum_{k = 1}^3 \tilde{F}_k(\bm{H})\,.
    \end{aligned}
    \end{equation}
In the second line above, we have plugged in the generators of the complete intersection of hyperplanes along which we blow up the $\SU(2)$ model, as described in \cref{eq:SU2blowupgen}. Likewise, we recall (\cref{eq:smoothpush}) that the pushforward of an arbitrary function of the class $\bm{H}$ is given by
    \begin{equation}
    \begin{aligned}
        \varpi_*(\sum_{k = 1}^3 \tilde{F}_k(\bm{H})) &=  \sum_{a = 1}^3 \frac{\sum_{k = 1}^3 \tilde{F}_k(-L_a)}{\prod_{b \ne a} (L_a -  L_b)} \\
        &=-\frac{1}{2} \frac{\sum_{k = 1}^3 \tilde{F}_k(2 K)}{K^2} + \frac{1}{3} \frac{\sum_{k = 1}^3 \tilde{F}_k(3 K)}{K^2} + \frac{1}{6} \frac{\sum_{k = 1}^3 \tilde{F}_k(0)}{K^2}\,,
    \end{aligned}
    \end{equation}
where on the right-hand side of the above equation we have imposed the CY condition $L = -K$. The computation of the pushforward of the generating function \cref{eq:SU2genfun} is now a straightforward (albeit messy) substitution of the function $e^{\hat{J}}$ for the dummy function $\tilde{F}(\bm{E}_1)$ in the above expressions. Ultimately, we find
    \begin{equation}
         Z_\phi = -e^{\phi^\alpha D_\alpha} \frac{-e^{K \phi^0} + \frac{2 K e^{\Sigma \phi^1} + \Sigma e^{-2 K \phi^1}}{2 K + \Sigma}}{K}\,.
    \end{equation}
Note that the change to the basis of Cartan divisors is trivial, namely $\alpha^1 = \phi^1$. We can cross-check this result by verifying \cref{eq:Cartan} explicitly:
    \begin{equation}
        \left.\left(\frac{\partial}{\partial \phi^1}\right)^2 Z_\phi\right|_{\phi^I = 0} = \left.-e^{\phi^\alpha D_\alpha} \frac{2 \Sigma \left(2 K e^{-2 K \phi^1} + \Sigma e^{\Sigma \phi_1}\right)}{2 K + \Sigma}\right|_{\phi^I = 0} = - 2 \Sigma\,.
    \end{equation}

\subsubsection{$\SO(2 r + 1)$ model}

Assume $r > 3$. The $\SO(2 r + 1)$ Tate model is characterized by a $\singtype{I}_{r - 3}^{* \text{non-split}}$ Kodaira singularity over a divisor $\Sigma \subset B^{(n - 1)}$. This model can be constructed via the Tate tuning
    \begin{equation}
        a_1 = a_{1, 1} \sigma\,, \quad a_2 = a_{2, 1} \sigma\,, \quad a_3 = a_{3, \floor*{\frac{r + 1}{2}}} \sigma^{\floor*{\frac{r + 1}{2}}}\,, \quad a_4 = a_{4, \ceil*{\frac{r + 1}{2}}} \sigma^{\ceil*{\frac{r + 1}{2}}}\,, \quad a_6 = a_{6, r} \sigma^r\,.
    \end{equation}
We consider the family of resolutions $X_r^{(n)} \to X_0^{(n)}$ given by the following sequence of blowups~\cite{Bhardwaj:2018yhy}:
    \begin{equation}
        \label{eq:Bres}
            X_r^{(n)} \xrightarrow{(e_{r - 2}, e_{r - 1} | e_r)} X_{r - 1}^{(n)} \xrightarrow{(*, e_{r - 2} | e_{r - 1})} \dotsb \to X_3^{(n)} \xrightarrow{(x, e_2 | e_3)} X_2^{(n)} \xrightarrow{(y, e_1 | e_2)} X_1^{(n)} \xrightarrow{(x, y, \sigma | e_1)} X_0^{(n)}\,,
    \end{equation}
where
    \begin{equation}
        * = \begin{cases}
            x, & r \text{ even} \\
            y, & r \text{ odd}
        \end{cases}\,.
    \end{equation}
We can expand the classes of the proper transforms of the divisors $e_i = 0$ as
    \begin{equation}
        \label{eq:SOOddlin}
        \bm{e}_i = e_i^j \bm{E}_j = \hat{e}_i^j \hat{\bm{D}}_j
    \end{equation}
and hence the change of basis from the $\bm{E}_i$ to the $\hat{\bm{D}}_i$ basis is given by
        \begin{equation}
            \alpha^i =  e_j^i (\hat{e}^{-1})_k^j \phi^k.
        \end{equation}
For $r \ge 5$, the matrices $e_i^j, \hat{e}_i^j$ are defined implicitly by the following representation of \cref{eq:SOOddlin}:
    \begin{equation}
        \begin{pmatrix}
            \bm{e}_1 \\
            \bm{e}_2 \\
            \bm{e}_3 \\
            \bm{e}_4 \\
            \vdots \\
            \bm{e}_{r - 2} \\
            \bm{e}_{r - 1} \\
            \bm{e}_r
        \end{pmatrix} =
        \begin{pmatrix}
            \bm{E}_1 - \bm{E}_2 \\
            \bm{E}_2 - \bm{E}_3 \\
            \bm{E}_3 - \bm{E}_4\\
            \bm{E}_4 - \bm{E}_5 \\
            \vdots \\
            \bm{E}_{r - 2} - \bm{E}_{r - 1} - \bm{E}_r \\
            \bm{E}_{r - 1} - \bm{E}_r \\
            \bm{E}_r
        \end{pmatrix} =
        \begin{pmatrix}
            \hat{\bm{D}}_2 \\
            \hat{\bm{D}}_1 + \hat{\bm{D}}_2 + \hat{\bm{D}}_3 \\
            \hat{\bm{D}}_3 + \hat{\bm{D}}_4 \\
            \hat{\bm{D}}_4 + \hat{\bm{D}}_5 \\
            \vdots \\
            \hat{\bm{D}}_{r - 2} \\
            \hat{\bm{D}}_r \\
            \hat{\bm{D}}_{r - 1}
        \end{pmatrix}\,.
    \end{equation}
For $r = 5$, the transformation is defined by
    \begin{equation}
        \begin{pmatrix}
            \bm{e}_1 \\
            \bm{e}_2 \\
            \bm{e}_3 \\
            \bm{e}_4 \\
            \bm{e}_5
        \end{pmatrix} =
        \begin{pmatrix}
            \bm{E}_1 - \bm{E}_2 \\
            \bm{E}_2 - \bm{E}_3 \\
            \bm{E}_3 - \bm{E}_4 - \bm{E}_5 \\
            \bm{E}_4 - \bm{E}_5 \\
            \bm{E}_5
        \end{pmatrix} =
        \begin{pmatrix}
            \hat{\bm{D}}_2 \\
            \hat{\bm{D}}_1 + \hat{\bm{D}}_2 + \hat{\bm{D}}_3 \\
            \hat{\bm{D}}_3 \\
            \hat{\bm{D}}_5 \\
            \hat{\bm{D}}_4
        \end{pmatrix}\,.
    \end{equation}
For $r = 4$, the transformation is defined by
    \begin{equation}
        \begin{pmatrix}
            \bm{e}_1 \\
            \bm{e}_2 \\
            \bm{e}_3 \\
            \bm{e}_4
        \end{pmatrix} =
        \begin{pmatrix}
            \bm{E}_1 - \bm{E}_2 \\
            \bm{E}_2 - \bm{E}_3 - \bm{E}_4 \\
            \bm{E}_3 - \bm{E}_4 \\
            \bm{E}_4
        \end{pmatrix} =
        \begin{pmatrix}
            \hat{\bm{D}}_2 \\
            \hat{\bm{D}}_1 + \hat{\bm{D}}_2 \\
            \hat{\bm{D}}_3 \\
            \hat{\bm{D}}_4
        \end{pmatrix}\,.
    \end{equation}
The case $r = 3$, namely the $\SO(7)$ model, is special. This model is characterized by a $\singtype{I}_0^{* \text{semi-split}}$ fiber. We consider the resolution defined by the composition of blowups
    \begin{equation}
        X_3^{(n)} \xrightarrow{(x, e_2 | e_3)} X_2^{(n)} \xrightarrow{(y, e_1 | e_2)} X_1^{(n)} \xrightarrow{(x, y, \sigma | e_1)} X_0^{(n)}\,.
    \end{equation}
The transformation to the basis of Cartan divisors is defined by
    \begin{equation}
        \begin{pmatrix}
            \bm{E}_1 - \bm{E}_2 \\
            \bm{E}_2 - \bm{E}_3 \\
            \bm{E}_3
        \end{pmatrix} =
        \begin{pmatrix}
            \hat{\bm{D}}_2 \\
            \hat{\bm{D}}_2 + \hat{\bm{D}}_3 \\
            \hat{\bm{D}}_1
        \end{pmatrix}\,.
    \end{equation}
For $r = 2, 1$, one can use the accidental (Lie algebra) isomorphisms $\SO(5) = \Sp(2), \SO(3) = \SU(2)$, and hence we omit these cases.

As an example, consider the case $r = 3$, i.e., the $\SO(7)$ model. After changing to the basis of Cartan divisors, i.e., $\alpha^0 = \phi^0, \alpha^1 = \phi^2 - \phi^3, \alpha^2 = -2 \phi^2 + 2 \phi^3, \alpha^3 = \phi^1 - \phi^3$, we obtain the following expression
    \begin{equation}
    \begin{aligned}
        Z_\phi &= \frac{e^{\phi^\alpha D_\alpha - K \phi^2}}{K (K + \Sigma) (2 K + \Sigma) (3 K + 2 \Sigma)} \Big[-K \Sigma (K + \Sigma) e^{-(2 K + \Sigma) (\phi^1- 2 \phi^2 + 2 \phi^3)} \\
        &\qquad - K \Sigma (2 K + \Sigma) e^{-(2 K + \Sigma) \phi^1 + K \phi^2 + 2 (K + \Sigma) \phi^3} \\
        &\qquad - \Sigma (3 K^2 + 5 K \Sigma + 2 \Sigma^2) - K (6 K^2 + 7 K \Sigma + 2 \Sigma^2) e^{\Sigma \phi^1 + K \phi^2} \\
        &\qquad + (6 K^3 + 13 K^2 \Sigma + 9 K \Sigma^2 + 2 \Sigma^3) e^{K (\phi^0 + \phi^2)}\Big]\,.
    \end{aligned}
    \end{equation}

\subsubsection{$\Sp(r)$ model}

Assume $r > 1$ (we omit the case $\Sp(1) = \SU(2)$). This model is characterized by a $\singtype{I}_{2 r + 1}^\text{non-split}$ Kodaira singularity over a divisor $\Sigma \subset B^{(n - 1)}$, and can be constructed via the Tate tuning
    \begin{equation}
        a_1 = 0\,, \quad a_2 = a_2\,, \quad a_3 = 0\,, \quad a_4 = a_{4, r + 1} \sigma^{r + 1}\,, \quad a_6 = a_{6, 2 r + 1} \sigma^{2 r + 1}\,.
    \end{equation}
We consider the family of resolutions $X^{(n)}_r \to X_0^{(n)}$ defined by the follow sequence of blowups~\cite{Esole:2017kyr}:
    \begin{equation}
        \label{eq:Cres}
        X_r^{(n)} \xrightarrow{(x, y, e_{r - 1} | e_r)} X_{r - 1}^{(n)} \xrightarrow{(x, y, e_{r - 2} | e_{r - 1})} \dotsb \xrightarrow{(x, y, e_1 | e_2)} X_1^{(n)} \xrightarrow{(x, y, \sigma | e_1)} X_0^{(n)}\,.
    \end{equation}
The change of basis from the $\bm{E}_i$ to the $\hat{\bm{D}}_i$ basis is given by
    \begin{equation}
        \alpha^1 = \phi^1\,, \quad \alpha^i = \phi^i - \phi^{i - 1}\,, \quad i = 2, \dotsc, r\,.
    \end{equation}
As an example, consider the case $r = 2$, i.e., the $\Sp(2)$ model. After changing to the basis of Cartan divisors, i.e., $\alpha^0 = \phi^0, \alpha^1 = \phi^1, \alpha^2 = -\phi^1 + \phi^2$, we obtain
    \begin{equation}
    \begin{aligned}
        Z_\phi &= \frac{e^{\phi^\alpha D_\alpha - 2 K \phi^1}}{K (K + \Sigma) (2 K + \Sigma)} \Big[(2 K^2 + 3 K \Sigma + \Sigma^2) e^{K (\phi^0 + 2 \phi^1)} - \Sigma (K + \Sigma) \\
        &\qquad - K \Sigma e^{(2 K + \Sigma) (2 \phi^1 - \phi^2)} - K (2 K + \Sigma) e^{2 K \phi^1 + \Sigma \phi^2}\Big]\,.
    \end{aligned}
    \end{equation}

\subsubsection{$\SO(2 r)$ model}

Assume $r > 4$. The $\SO(2 r)$ Tate model is characterized by a $\singtype{I}_{r - 4}^{* \text{split}}$ Kodaira singularity over a divisor $\Sigma \in B^{(n - 1)}$, and can be constructed via the Tate tuning\footnote{The tuning described in \cref{eq:SO2rTate} may not describe the most general F-theory Tate model that exhibits $\SO(2 r)$ gauge symmetry; see \cref{tab:Tatetable} for more general constructions. Despite this, we nevertheless expect that the sequence of blowups in \labelcref{eq:Dres} will also resolve the more general $\SO(2 r)$ models, and hence that the intersection numbers of divisors are indistinguishable.}
    \begin{equation}
    \begin{gathered}
        \label{eq:SO2rTate}
        a_1 = a_{1, 1} \sigma\,, \quad a_2 = a_{2, 1} \sigma\,, \quad a_3 = a_{3, \ceil*{\frac{r - 1}{2}}} \sigma^{\ceil*{\frac{r - 1}{2}}}\,, \\
        a_4 = a_{4, \floor*{\frac{r + 1}{2}}} \sigma^{\floor*{\frac{r + 1}{2}}}\,, \quad a_6 = a_{6, r - \frac{1 + (-1)^r}{2}} \sigma^{r - \frac{1 + (-1)^r}{2}}\,.
    \end{gathered}
    \end{equation}
When $r$ is even, one must impose the additional condition that
    \begin{equation}
        \left.\frac{a_4^2 - 4 a_2 a_6}{\sigma^r}\right|_{\sigma = 0}
    \end{equation}
is a perfect square to ensure that the Kodaira fiber is split.\footnote{In practice, it is convenient to impose this condition by setting $a_6 = 0$, although the singular elliptic fibrations defined by this choice of parameter do not represent the most general class of $\SO(2 r)$ models.} We consider the family of resolutions $X^{(n)}_r \to X_0^{(n)}$ defined by the following sequence of blowups~\cite{Bhardwaj:2018yhy}:
    \begin{equation}
        \label{eq:Dres}
        X_r^{(n)} \xrightarrow{(e_{r - 3}, e_{r - 2} | e_r)} X_{r - 1}^{(n)} \xrightarrow{(*, e_{r - 2} | e_{r - 1})} \dotsb \to X_3^{(n)} \xrightarrow{(x, e_2 | e_3)} X_2^{(n)} \xrightarrow{(y e_1 | e_2)} X_1^{(n)} \xrightarrow{(x, y, \sigma | e_1)} X_0^{(n)}\,,
    \end{equation}
where
    \begin{equation}
        * = \begin{cases}
            x, & r \text{ even} \\
            y, & r \text{ odd}
        \end{cases}\,.
    \end{equation}
We can expand the classes of the proper transforms of the divisors $e_i = 0$ as
    \begin{equation}
        \label{eq:SOEvenlin}
        \bm{e}_i = e_i^j \bm{E}_j = \hat{e}_i^j \hat{\bm{D}}_j
    \end{equation}
and hence
    \begin{equation}
        \alpha^i = e_j^i (\hat{e}^{-1})_k^j \phi^k\,.
    \end{equation}
For $r \ge 6$, the matrices $e_i^j, \hat{e}_i^j$ are defined implicitly by the following representation of \cref{eq:SOEvenlin}:
    \begin{equation}
        \begin{pmatrix}
            \bm{e}_1 \\
            \bm{e}_2 \\
            \bm{e}_3 \\
            \bm{e}_4 \\
            \vdots \\
            \bm{e}_{r - 3} \\
            \bm{e}_{r - 2} \\
            \bm{e}_{r - 1} \\
            \bm{e}_r
        \end{pmatrix} =
        \begin{pmatrix}
            \bm{E}_1 - \bm{E}_2 \\
            \bm{E}_2 - \bm{E}_3 \\
            \bm{E}_3 - \bm{E}_4 \\
            \bm{E}_4 - \bm{E}_5 \\
            \vdots \\
            \bm{E}_{r - 3} - \bm{E}_{r - 2} -\bm{E}_r \\
            \bm{E}_{r - 2} - \bm{E}_{r - 1} - \bm{E}_r \\
            \bm{E}_{r - 1} \\
            \bm{E}_r
        \end{pmatrix} =
        \begin{pmatrix}
            \hat{\bm{D}}_2 \\
            \hat{\bm{D}}_1 + \hat{\bm{D}}_2 + \hat{\bm{D}}_3 \\
            \hat{\bm{D}}_3 + \hat{\bm{D}}_4 \\
            \hat{\bm{D}}_4 + \hat{\bm{D}}_5 \\
            \vdots \\
            \hat{\bm{D}}_{r - 3} \\
            \hat{\bm{D}}_r \\
            \hat{\bm{D}}_{r - 1} \\
            \hat{\bm{D}}_{r - 2}
        \end{pmatrix}\,.
    \end{equation}
For $r = 5$, the transformation is defined by
    \begin{equation}
        \begin{pmatrix}
            \bm{e}_1 \\
            \bm{e}_2 \\
            \bm{e}_3 \\
            \bm{e}_4 \\
            \bm{e}_5
        \end{pmatrix} =
        \begin{pmatrix}
            \bm{E}_1 - \bm{E}_2 \\
            \bm{E}_2 - \bm{E}_3 - \bm{E}_5 \\
            \bm{E}_3 - \bm{E}_4 - \bm{E}_5 \\
            \bm{E}_4 \\
            \bm{E}_5
        \end{pmatrix} =
        \begin{pmatrix} \hat{\bm{D}}_2 \\
            \hat{\bm{D}}_1 + \hat{\bm{D}}_2 \\
            \hat{\bm{D}}_5 \\
            \hat{\bm{D}}_4 \\
            \hat{\bm{D}}_3
        \end{pmatrix}\,.
    \end{equation}
The case $r = 4$, namely the $\SO(8)$ model, is special. Here, we must impose the condition that
    \begin{equation}
        \left.\frac{a_2^2 - 4 a_4}{\sigma^4}\right|_{\sigma = 0}
    \end{equation}
is a perfect square; formally, we write
    \begin{equation}
        \alpha^2 \equiv 4 a_{4, 2} - a_{2, 1}^2
    \end{equation}
and consider the resolution defined by the following sequence of blowups
    \begin{equation}
        X_4^{(n)} \xrightarrow{(\sigma \alpha_2^\pm z + 2 e_3 x, e_2 | e_4)} X_3^{(n)} \xrightarrow{(x, e_2 | e_3)} X_2^{(n)} \xrightarrow{(y, e_1 | e_2)} X_1^{(n)} \xrightarrow{(x, y, \sigma | e_1)} X_0^{(n)}\,,
    \end{equation}
where $\alpha_2^\pm \equiv \alpha \pm a_{2, 1}$. The transformation to the Cartan divisor basis is defined by by
    \begin{equation}
        \begin{pmatrix}
            \bm{e}_1 \\
            \bm{e}_2 \\
            \bm{e}_3 \\
            \bm{e}_4
        \end{pmatrix} =
        \begin{pmatrix}
            \bm{E}_1 - \bm{E}_2 \\
            \bm{E}_2 - \bm{E}_3 - \bm{E}_4 \\
            \bm{E}_3 \\
            \bm{E}_4
        \end{pmatrix} =
        \begin{pmatrix}
            \hat{\bm{D}}_2 \\
            \hat{\bm{D}}_1 + \hat{\bm{D}}_2 \\
            \hat{\bm{D}}_4 \\
            \hat{\bm{D}}_3
        \end{pmatrix}\,.
    \end{equation}
For $r = 3, 2$, one can use the accidental (Lie algebra) isomorphisms $\SO(6) = \SU(4), \SO(4) = \SU(2) \times \SU(2)$, and hence we omit these cases.

As an example, consider the case $r = 4$, i.e., the $\SO(8)$ model. After changing to the basis of Cartan divisors, i.e., $\alpha^0 = \phi^0, \alpha^1 = -\phi^1 + \phi^2 , \alpha^2 = 2 \phi^1 - \phi^2, \alpha^3 = -\phi^1 + \phi^3, \alpha^4 = - \phi^1 + \phi^4$, we obtain the following generating function:
    \begin{equation}
        \begin{aligned}
            Z_\phi &= \frac{e^{\phi^\alpha D_\alpha - K \phi^2}}{K (K + \Sigma) (2 K + \Sigma) (3 K + 2 \Sigma) (4 K + 3 \Sigma)}\\
            &\qquad \times \Big[-K \Sigma (2 K^2 + 3 K \Sigma + \Sigma^2) e^{(4 K + 3 \Sigma) \phi^1 + K \phi^2 - (2 K + \Sigma) (\phi^3 + \phi^4)} \\
            &\qquad\qquad - K \Sigma (6 K^2 + 7 K \Sigma + 2 \Sigma^2) e^{K \phi^2 - (2 K + \Sigma) \phi^3 + 2 (K + \Sigma) \phi^4} \\
            &\qquad\qquad - K \Sigma  (4 K^2 + 7 K \Sigma + 3 \Sigma^2) e^{ -(2 K + \Sigma) (\phi^1 - 2 \phi^2 + \phi^3 + \phi^4)} \\
            &\qquad\qquad - \Sigma  (12 K^3 + 29 K^2 \Sigma + 23 K \Sigma^2 + 6 \Sigma^3) \\
            &\qquad\qquad - K (24 K^3 + 46 K^2 \Sigma + 29 K \Sigma^2 + 6 \Sigma^3) e^{K \phi^2 + \Sigma \phi^3}\\
            &\qquad\qquad + (24 K^4 +  70 K^3 \Sigma + 75 K^2 \Sigma^2 +35 K \Sigma^3 + 6 \Sigma^4) e^{K (\phi^0 + \phi^2)}\Big]\,.
        \end{aligned}
    \end{equation}

\section{Discussion and future directions}
\label{sec:discussion}

We have presented an efficient algorithm for computing topological intersection numbers of divisors $\hat{D}_I$ in a resolved elliptic fibration $X^{(n)}$ defined over an arbitrary smooth projective base of complex dimension $n - 1$, where the explicit computation of intersection numbers is carried out by using the methods of~\cite{Esole:2017kyr} to compute their pushforward to the Chow ring of the base and evaluating the intersection numbers in terms of specific intersection products of divisors of the base. Although we have focused in particular on a subset of cases in which $X^{(n)} \to X_0^{(n)}$ resolves a singular hypersurface $X_0^{(n)}$ of a smooth projective bundle, and for which the resolution $X^{(n)} \to X_0^{(n)}$ is comprised of a sequence of blowups along complete intersections of hyperplanes in the ambient space, the methods we describe can be generalized to complete intersection Calabi--Yau (CICY) varieties and genus one fibrations. We have applied our algorithm to the specific case of F-theory Tate models with classical gauge groups by building a \emph{Mathematica} package that explicitly evaluates the pushforwards of generating functions of intersection numbers, focusing on the exponential of the K\"ahler class as an optimal choice of generating function. We have used this \emph{Mathematica} package to explicitly compute the generating functions for all F-theory Tate models with gauge groups given by simple classical Lie groups with rank up to 20; the results of these computations are collected in an ancillary \emph{Mathematica} notebook accompanying the upload of this preprint to the arXiv.

Our algorithm is a clear adaptation of the strategy utilized in~\cite{Esole:2017kyr,Esole:2018tuz,Esole:2018bmf} to compute Euler characteristics and other characteristic numbers of smooth and resolved elliptic fibrations---namely, we organize the topological intersection numbers into an analytic generating function and attempt to optimize the symbolic manipulations needed to evaluate the action of the pushforward map on this generating function. A key part of this optimization is to use the pushforward maps to pre-define a set of dummy functions taking the symbolic form of the image of an arbitrary function under $m \le r + 1$ ($r$ is the rank of the gauge group) pushforward maps. After defining this ``tower'' of dummy functions, the final step of the algorithm is to substitute the actual generating function for the first dummy function, which sits at the ``top'' of the tower, and use this generating function to recursively define the remaining pushforwards explicitly via direct substitution. The primary computational expense appears to arise from performing the symbolic substitutions need to evaluate the pushforward maps on a specific choice of function. In particular, these substitutions appear to introduce various spurious poles that need to be canceled in order to complete the computation. We use the \emph{Mathematica} functions \texttt{Apart[]} and \texttt{Cancel[]} to eliminate these spurious poles. Looking ahead, it would be interesting and potentially fruitful to try and optimize this specific aspect of the computation.

Regarding computational expense, one apparent advantage of using the exponential of the K\"ahler class as a generating function is the fact that the exponential is a succinct symbolic expression without any spurious poles. In practice, we have found that the function \texttt{push[]} in the \emph{Mathematica} package \texttt{IntersectionNumbers.m} can compute the pushforward of the exponential of the K\"ahler class (sometimes significantly) faster than the pushforward of other characteristic classes, e.g., the total Chern class. In fact, it appears that the fastest way to compute most characteristic classes of fixed degree is to first compute the generating function of intersection numbers and then use the generating function to substitute the intersection numbers into the explicit expression for the characteristic class. For example, in generic cases, the fastest method to compute Chern classes $c_k(X^{(n)})$ is to extract them from a power series expansion of the total Chern class,
    \begin{equation*}
        c_\varepsilon(X^{(n)}) = 1 + c_1(X^{(n)}) \varepsilon + c_2(X^{(n)}) \varepsilon^2 + \dotsb\,,
    \end{equation*}
and then substitute intersection numbers extracted from the pushforward of the generating function into the degree-$k$ monomials of $c_k(X^{(n)})$.

The creation of this algorithm was motivated by a need to compute characteristic numbers for F-theory Tate models with non-abelian gauge groups of large rank. To the authors' knowledge, this algorithm represents a significant reduction in the computational expense required to evaluate intersection numbers using the methods of~\cite{Esole:2017kyr}, and has potential applications to various problems in string theory. We briefly discuss a few potential future directions:
    \begin{itemize}
        \item
            It has been conjectured that only a finite number of Calabi--Yau compactifications exist~\cite{YauReview} for given dimension and supersymmetry. Motivated by this conjecture, swampland constraints on the landscape of consistent supergravity theories are commonly presented as numerical bounds on kinematic data, for example, as upper bounds on the ranks of gauge groups or numbers of matter multiplets transforming in a given subset of admissible representations---see, e.g.,~\cite{Tarazi:2021duw}. The computational algorithm of this paper could be used to more easily explore swampland-like upper bounds on kinematic data prohibiting F-theory vacua with large gauge groups, in particular by studying the consistency of supersymmetry-protected data characterizing the low-energy theory that is encoded in intersection numbers.
        \item
            The profusion of topologically distinct resolutions that exist for singular CY varieties makes any large-scale effort to explore the landscape of smooth CY phases in terms of topological invariants like intersection numbers a challenging and computationally expensive endeavor. In order to reduce this computational expense, it would be preferable to be able to study equivalence classes of smooth CY that belong to the same extended K\"ahler cone, while still retaining control over geometric data that determine the kinematics of the low-energy effective theories defined by these CY varieties. For this kind of approach to work, it is necessary to have a clear understanding of topological invariants of CY varieties that are also, in a precise sense, independent of the choice of (crepant) resolution. At present, it seems that such ``resolution invariants'' are not thoroughly understood for CY $n$-folds with $n \ge 3$. F-theory provides compelling evidence that singular \emph{elliptic} CY varieties, at least, can be characterized by a rich set of resolution invariants intimately related to the low-energy gauge theory defined by the compactification. One set of potential invariants are the intersection pairing matrices
                \begin{equation}
                    Q_{(I_1 \dotsm I_k) (J_1 \dotsm J_l)} = \left.\left(\frac{\partial}{\partial\phi^{I_1}} \dotsm \frac{\partial}{\partial\phi^{I_k}}\right) \left(\frac{\partial}{\partial\phi^{J_1}}   \dotsm \frac{\partial}{\partial\phi^{J_l}}\right) Z_\phi\right|_{\phi = 0}\,, \quad k + l = n\,.
                \end{equation}
            For example, it was conjectured in~\cite{Jefferson:2021bid} that the intersection pairing of vertical four-cycles in smooth elliptic CY $4$-folds defines a resolution-invariant lattice, and it is tempting to further speculate that this invariance extends to all pairing matrices for arbitrary smooth elliptic CY $n$-folds, though we stress that this assertion is (to the authors' knowledge) unproven and moreover has not been thoroughly explored, at least in the string theory literature. The pairing matrices $Q$ are very naturally expressed in terms of intersection products of divisors in $B^{(n - 1)}$ by means of the pushforward formulas described in this paper, and hence the pushforwards of generating functions we compute could be used to explore the potential resolution invariance of the pairing matrices $Q$ in a large numbers of examples.
        \item
            We also remark that it was shown in~\cite{Huang:2018esr} that the majority of known CY 3-folds (and possibly also CY 4-folds) admit at least one phase with an elliptic or genus one fibration, and hence define suitable backgrounds for F-theory compactifications. The pervasiveness of elliptic CY $n$-folds among the set of known CY $n$-folds in turn suggests that most (known) CY $n$-folds admit an F-theory description, or somewhat more speculatively, that most known CY $n$-folds can be characterized by a collection of birational invariants presumably expressed in terms of local geometric data and gauge theoretic structures describing the low-energy effective description of the F-theory compactification, such as the pairing matrices $Q$ described above. Identifying a list of birational invariants for CY $n$-folds of a given dimension $n$ could potentially lead to a simpler method to map out the landscape of CYs that in principle does not require the computation of resolutions. The ``experimental'' identification of birational invariants that are constructed from intersection numbers of divisors could be greatly facilitated by using the pushforwards of generating functions described in this paper, as the pushforward map guarantees that the intersection numbers are expressed in terms of special divisor classes of $B^{(n - 1)}$, which remain well-defined in the singular F-theory limit.
    \end{itemize}
We hope to use the methods developed in this paper to explore these and related ideas at the interface of string theory and algebraic geometry/intersection theory in the near future.

\section*{Acknowledgements}
We are extremely grateful to Paolo Aluffi, Mboyo Esole, and Monica Kang for many illuminating discussions about the foundational material upon which this work is based, along with Manki Kim, Ling Lin, Fabian Ruehle, Houri-Christina Tarazi, and Washington Taylor. We thank Manki Kim and Washington Taylor on valuable comments on earlier versions of this paper. PJ was supported by DOE grant DE-SC00012567. APT was supported by DOE (HEP) Award DE-SC0013528.

\appendix

\section{Notation}
\label{sec:notation}
Below is a list of notation commonly used throughout this document. Note that we abuse notation by using the same symbol to denote a divisor and its class in the Chow ring. Moreover, when a divisor class is the pullback of a divisor class belonging to a different Chow ring we do not explicitly indicate the pullback map when the Chow ring to which the divisor class in question belongs is otherwise clear from the context, e.g., we use the symbol $\bm{E}_{i - 1}$ to denote the class $f_i^*(\bm{E}_{i - 1})$ in the Chow ring of $Y_i^{(n + 1)}$.
    \begin{itemize}
        \item $\boxed{a_n}$: Sections of the anticanonical class, i.e., $[a_n = 0] = n(-K)$. When the sections are tuned to vanish over a gauge divisor $\sigma = 0$, we write $a_n = a_{n, m_n} \sigma^m$ with $[a_{n, m_n} = 0] = n(-K) - m_n \Sigma$.
        \item $\boxed{B^{(n - 1)}}$: Smooth K\"ahler base of the singular elliptic CY $n$-fold, $X_0^{(n)} \to B^{(n - 1)}$.
        \item $\boxed{D_\alpha}$: Basis of divisors of $B$.
        \item $\boxed{D D'}$: Intersection product of pair of divisors whose classes are, by abuse of notation, also denoted $D, D'$.
        \item $\boxed{\hat{D}_I}$: Basis of divisors in $X^{(n)}$, where the ``hat'' decoration distinguishes divisors in $X^{(n)}$ from divisors in $B^{(n - 1)}$. For smooth hypersurfaces that have the structure of a fibration $X^{(n)} \to B^{(n - 1)}$, this basis of divisors can be further partitioned as $\hat{D}_I = \hat{D}_\alpha, \hat{D}_{\hat{I}}$, where $\hat{D}_\alpha$ are pullbacks of divisors in $B^{(n - 1)}$ and $\hat{D}_{\hat{I}}$ cannot be written as pullbacks. Specializing further, for elliptic $n$-folds resolving singular $n$-folds with gauge group $\G$, the indices $I = 0, \alpha, i$ label (resp.) the zero section $\hat{D}_0$, pullbacks of divisors $\hat{D}_\alpha = \pi^* D_\alpha$ in the base $B^{(n - 1)}$, and Cartan divisors $\hat{D}_i$.
        \item $\boxed{\hat{\bm{D}}_I}$: Divisor class in the ambient $(n + 1)$-fold $Y^{(n + 1)}$ whose restriction to the class $\bm{X}^{(n)}$ of the hypersurface $X^{(n)}$ in the Chow ring of $Y^{(n + 1)}$ is the class of a divisor in $X^{(n)}$, i.e., $\hat{D}_I = \hat{\bm{D}}_I \cap \bm{X}^{(n)}$.
        \item $\boxed{{\bm{D}}}$: Divisors of the ambient projective bundle $Y^{(n + 1)}$, e.g., the hyperplane class $\bm{H}$. A basis of divisors of $Y^{(n + 1)}$ includes as a special subset the divisors $\bm{D}_\alpha$, which are the pullbacks of divisors (with respect to the lift of the projection map $\pi_*\colon X^{(n)} \to B^{(n - 1)}$) $D_\alpha \subset B^{(n - 1)}$.
        \item $\boxed{\Delta}$: Discriminant of the Weierstrass equation. The locus $\Delta = 0$ in $B^{(n - 1)}$ is the discriminant locus, over which the elliptic fibers of $X_0^{(n)}$ develop singularities.
        \item $\boxed{e_i}$: Local coordinate whose zero locus $e_i = 0$ in $Y^{(n + 1)}$ is (the proper transform of) the exceptional divisor $\bm{E}_i$.
        \item $\boxed{e^{\hat{J}}}$: Generating function of intersection numbers. A central result of this paper is an efficient algorithm for evaluating the pushforward of $e^{\hat{J}}$ to the Chow ring of $B^{(n - 1)}$.
        \item $\boxed{\G}$: F-theory gauge symmetry group encoded in the singularities of $X_0^{(n)}$ over the codimension-one locus $\Sigma \subset B^{(n - 1)}$. When $\G$ consists of multiple simple factors $\G_s$, we take each factor to correspond to a codimension-one singularity over $\Sigma_s \subset B^{(n - 1)}$.
        \item $\boxed{\hat{J}}$: K\"ahler class $\hat{J} = \phi^I \hat{D}_I$.
        \item $\boxed{K}$: Canonical divisor class of $B^{(n - 1)}$, $K = K^\alpha D_\alpha$.
        \item $\boxed{\kappa_{i j}}$: Matrix elements of the inverse metric tensor $\kappa$ of a Lie algebra.
        \item $\boxed{\PD(\hat{D})}$: Poincar\'{e} dual of $\hat{D}$.
        \item $\boxed{\pi}$: Canonical projection map from the smooth $n$-fold to the base, $\pi\colon X^{(n)} \to B^{(n - 1)}$.
        \item $\boxed{\varpi}$: Canonical projection map from the ambient projective $(n + 1)$-fold to the base, $\varpi\colon Y^{(n + 1)} \to B^{(n - 1)}$.
        \item $\boxed{\Sigma}$: The divisor class of the codimension-one locus $\sigma = 0$ in the base supporting the simple gauge algebra $\Lie(\G)$, i.e., $\Sigma = \Sigma_s^\alpha D_\alpha$.
        \item $\boxed{X^{(n)}}$: Smooth elliptic $n$-fold $X^{(n)} \to X_0^{(n)}$ resolving the singular $n$-fold $X_0^{(n)}$. When the smooth $n$-fold is the result of an explicit finite sequence of blowups, we write $X_i^{(n)}$ to denote the proper transform of $X_0$ under the $i$th blowup.
        \item $\boxed{X_0^{(n)}}$: Singular elliptic Calabi--Yau (CY) $n$-fold (i.e., complex dimension $n$) defining the compactification space of a $(12 - n)$D F-theory model.
        \item $\boxed{Y^{(n + 1)}}$: Ambient $(n + 1)$-fold bundle in which the resolution $X^{(n)}$ is realized as a hypersurface, i.e., $X^{(n)} \subset Y^{(n + 1)}$. In practice, $Y^{(n + 1)} \equiv Y_r^{(n + 1)}$ is the image of the ambient projective bundle $Y_0^{(n + 1)}$ under a composition of $r$ blowups, with corresponding blowdown maps $f_i : Y_{i}^{(n + 1)} \to Y_{i - 1}^{(n + 1)}$.
        \item $\boxed{Y_0^{(n + 1)}}$: Ambient $(n + 1)$-fold projective bundle (i.e., a bundle over the base $B^{(n - 1)}$ in which the fibers are projective spaces) in which the singular CY $n$-fold is realized as a hypersurface, $X_0^{(n)} \subset Y_0^{(n + 1)}$.
        \item $\boxed{Z_\phi}$: We write the pushforward of the generating function of intersection numbers as $Z_\phi(L_a, \Sigma_s) = \pi_*(e^{\hat{J}})$, where $L_a, \Sigma_s$ are the characteristic divisor classes defining the singular elliptic $n$-fold $X_0^{(n)}$.
        \end{itemize}

\bibliographystyle{JHEP}
\bibliography{references}

\end{document}